\documentclass[%
  aps,
  prl,
  twocolumn,
  superscriptaddress,
  nofootinbib,
  longbibliography,
  a4paper 
]{revtex4-2}
\usepackage[english]{babel}
\usepackage[T1]{fontenc}
\usepackage[utf8]{inputenc} 
\usepackage{microtype}
\usepackage{graphicx}
\usepackage{amsmath,amssymb}
\usepackage{booktabs}
\usepackage{siunitx}
\usepackage{xcolor}

\usepackage{bm}
\usepackage{dcolumn}
\usepackage{xcolor}
\usepackage[version=4]{mhchem}

\definecolor{darkyellow}{RGB}{180,140,0}
\usepackage[colorlinks=true,allcolors=blue]{hyperref} 
\begin{document}

\title{RF Heating in a TESLA 9-Cell Cavity Using Modes of the First Dipole Passband: The reversed role of HOM Couplers}
\author{R. Schöder}     
\affiliation{Helmholtz-Zentrum Berlin für Materialien und Energie, Albert-Einstein-Str. 15, 12489 Berlin, Germany}
\affiliation{Technical University Dortmund, Dortmund, Germany}
\author{H.-W. Glock}    
\affiliation{Helmholtz-Zentrum Berlin für Materialien und Energie, Albert-Einstein-Str. 15, 12489 Berlin, Germany}
\author{J. Knobloch}
\affiliation{Helmholtz-Zentrum Berlin für Materialien und Energie, Albert-Einstein-Str. 15, 12489 Berlin, Germany}
\affiliation{University of Siegen, Siegen, Germany}
\author{A. Velez} 
\affiliation{Helmholtz-Zentrum Berlin für Materialien und Energie, Albert-Einstein-Str. 15, 12489 Berlin, Germany}
\affiliation{Technical University Dortmund, Dortmund, Germany}
\date{\today}

\begin{abstract}
Recent studies have demonstrated the possibility to increase the performance of bulk niobium cavities by baking under vacuum conditions, typically in a dedicated oven or with external ohmic heaters. 
Alternatively, the authors have reported on the possibilities and main limitations, mostly imposed by a poor coupling in warm, of transmitting rf power through the main fundamental power coupler to perform such treatments. 
In this work, the strong coupling of the higher-order-mode (HOM) antennas to modes above the fundamental passband is studied and presented as a way to maximize power transfer into the cavity, opposite to its standard role as energy dampers. 
The first dipole passband of a standard 9-cell TESLA-type 1.3~GHz SRF cavity is used, which could seem as a counter-intuitive choice in view of the dipole nature of the field distribution and the aim of an ideally homogeneous temperature distribution. 
The band choice, on the other hand, takes advantage of the beam-tube cut-off frequencies to avoid power losses by outward propagating fields.
The present study shows the experimental work done with a set of modes of the first dipole band modes in order to study the temperature homogeneity they generate with the help of an infrared camera. 
The presence of conductive and convective losses in the surrounding air significantly reduced heating rates and maximum temperatures compared with the situation under vacuum, but allowed for a much more detailed analysis of temperature gradients as compared to the use of a few discrete temperature sensors. 
Those gradients are mainly defined by the thermal conductivity and the distribution of the rf power deposition, thus (in a first-order consideration of temperature-independent thermal and electric conductivity) being only weakly dependent on absolute temperature levels which therefore makes them accessible under the aforementioned conditions. 
In the study we found both an almost ideal rf coupling, a power handling of a standard HOM coupler above 200~W cw and temperature spans on the cavity surface not exceeding 14~K.

\end{abstract}
\maketitle

\section{Introduction}
Thermal treatments of superconducting radio-frequency (SRF) cavities are established methods for improving their intrinsic quality factor $Q_0$ \cite{PhysRevApplied.13.014024,ito:srf2021-frofdv01,Goedecke_2026}. 
In-situ baking approaches, in which the cavity is not exposed to the atmosphere between thermal treatment and subsequent cryogenic operation, have also shown promising results \cite{Tamashevich2025,prudnikava:srf2025-tua04}. 
The strong temperature dependence of the process is also emphasized in this study.

As part of ongoing efforts to explore rf-driven in-situ processing techniques, previous studies by the authors \cite{Glock2024} and others \cite{goedecke:srf2025-thp63} have investigated the feasibility of heating TESLA cavities using modes of the fundamental passband. 
In particular, rf heating through the fundamental power coupler (FPC) was examined \cite{wk7y-8n6d,schoeder_ipac2026-mop7025}. 
However, this approach is limited by the coupling strength observed at room temperature, as the available adjustment range of the coupler does not necessarily provide sufficiently strong coupling for efficient rf power transfer.

The present work investigates an alternative approach based on rf heating through the existing higher-order-mode (HOM) couplers of a TESLA cavity, using resonances of the first TE-like dipole passband (\cite{Wanzenberg:415327}, p. 33). 
These modes, located in the frequency range of approximately 1.6--1.8~GHz, offer much more favorable coupling conditions, partially close to matching, when using the HOM couplers. 
Furthermore, the investigated frequency range lies well below the first beam-pipe cutoff frequency (2.25~GHz for the TE${11}$ mode) \cite{Wanzenberg:415327},
which (almost fully) prevents the propagation of rf power through the beam pipes. 
The modes of this TE passband nevertheless are characterized by higher wall current densities on the inter-cell irises as compared to the cell´s equator region. 
This fact combined with the intrinsic dipolar field distribution on the surfaces rises the question  whether the use of such modes would lead to a temperature homogeneity over the cavity's inner surface sufficient for future baking treatments. 
The study described paper therefore tests rf power coupling through HOM antennas while simultaneously studying the heat deposition profile and temperature homogeneity. 

Such a task requires a finer spatial temperature sampling than possible with even an extensive amount of discrete temperature sensors. 
Therefore it was decided to make use of infrared imaging, giving access to a large part of the outer cavity surface with a mm-scale resolution. 
In order to overcome partial shadowing, the cavity support and cabling were modified to allow for a partial rotation around the cavity axis during operation without compromising any rf connection. 
This gave visual access to more than 50\% of the outer cavity surface by executing a forth-pause-back rotation in certain time intervals, typically 60~s with 5~s of pause duration. 

Infrared thermometry on shiny metal surfaces is notoriously prone to significant uncertainties due to low emission factors. This has a strong impact on the temperature readouts that may heavily depend on surface orientation or local modifications of roughness. 
To overcome this restriction a carbon black coating was applied on the outer cavity surface. 
This enhances the radiative loss by a factor >~3, which reduces the absolute temperature level increase after stabilization for a fixed power deposition (i.e. level and field pattern) also by such a significant factor. 

\maketitle
\section{EXPERIMENTAL SETUP}

\begin{figure}[t]
  \centering
  \includegraphics[width=0.47\textwidth]{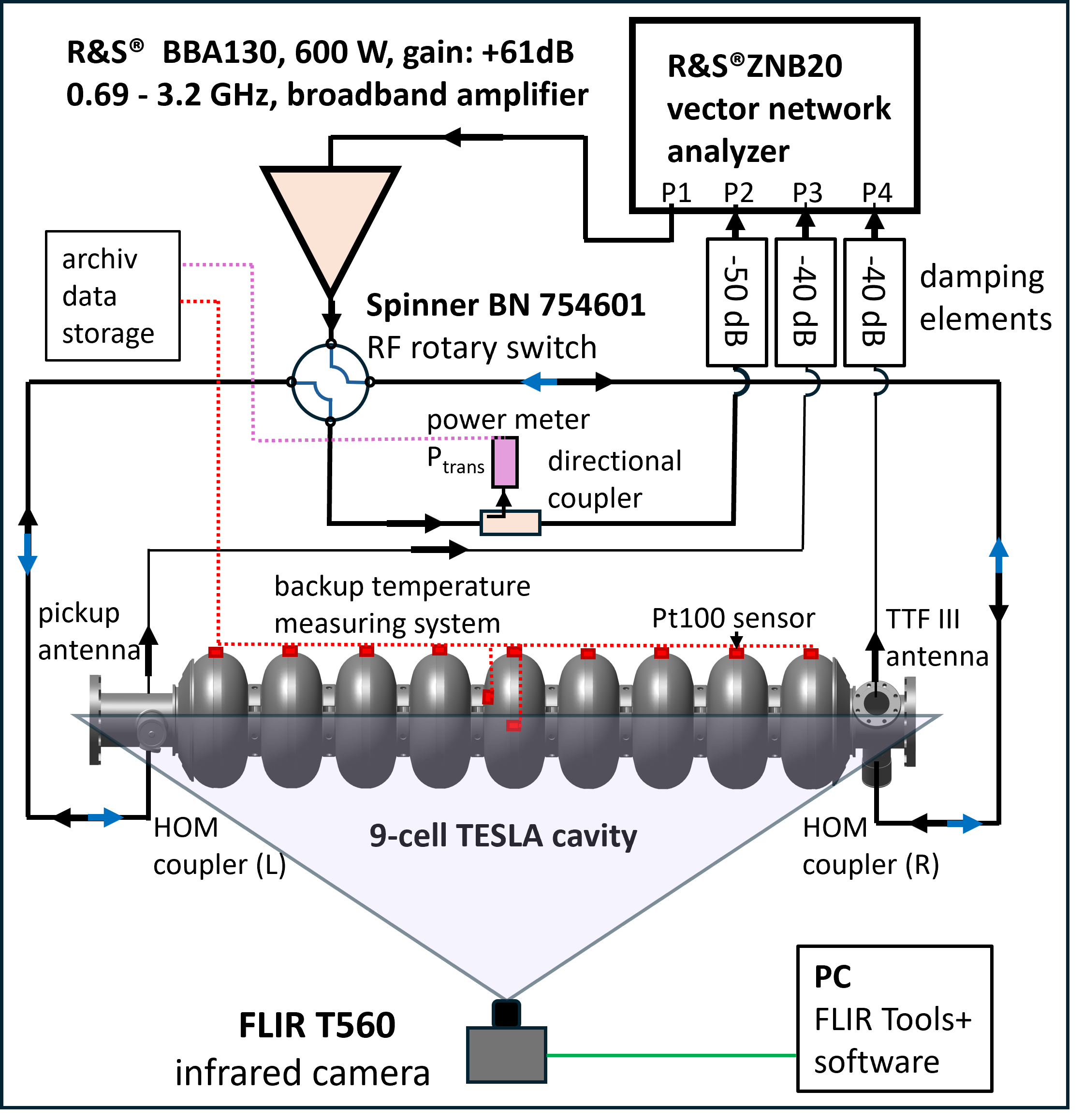} 
  \caption{Schematic of the experimental setup consisting of the rf heating system in an S21-like transmission configuration, including the R\&S\textsuperscript{\textregistered} ZNB20 vector network analyzer and BBA130 amplifier, the backup temperature measurement system using PT100 sensors, and the FLIR T560 infrared thermography system.}
  \label{fig:1}
\end{figure}

As it is being introduced the experimental goal of this work is to drive power into a 9-cell SRF cavity via the HOM antennas while monitoring the surface temperature distribution. 
To this end, an undressed niobium fine grade (RRR300) 1.3 GHz TESLA 9-cell cavity has been used. 
This cavity was used in previous studies on rf heating and results published in \cite{Glock2024, wk7y-8n6d,schoeder_ipac2026-mop7025}. 
The measurements were performed using a tabletop setup. No flanges were installed, and the cavity was fully exposed to ambient air.
The experimental setup consists of three main systems (cf.~Fig.~\ref{fig:1}): the rf power and measurement system required to deliver and read the forward and transmitted signals, an infrared thermography system to measure the surface temperature distribution of the cavity, and a sensor-based temperature measurement system used to complement and calibrate the infrared thermographic measurements. 

The core of the rf heating and diagnostic system consists of a Rohde \& Schwarz\textsuperscript{\textregistered} ZNB20 vector network analyzer (VNA) combined with a Rohde \& Schwarz\textsuperscript{\textregistered} BBA130 broadband amplifier. 
The VNA drives the amplifier via port P1, which operates over a frequency range of 0.69--3.2~GHz and delivers up to 600~W of rf power. 
A four-port Spinner BN 754601 rf rotary switch allows the amplified rf power to be fed either into the left (L) or the right (R) HOM coupler of the TESLA cavity. 
The respective opposite HOM coupler extracts the transmitted rf power, which is routed via the rotary switch to a bidirectional coupler. This is monitored by a power meter connected to the archive data storage. 
The through port of the directional coupler is connected back to VNA port P2 through a 50 dB rf attenuator. 
In addition, the transmitted powers captured by the pickup antenna and the TTF coupler antenna (consisting of an original cold part and a customized N-connector adaptor replacing its warm part) are routed through protective 40-dB-attenuators to ports P3 and P4 of the VNA.

For direct surface-temperature measurements, 11 PT100 sensors were used, with one sensor positioned at the equator of each cavity cell at the cavity's back side, thus not perturbing the infrared picture.
Two additional sensors were placed on the central cell, one of them within the infrared camera's view field allowing for a direct comparison of readout values of either system. 
The PT100 sensors were mounted on copper blocks, which were thermally coupled to the cavity surface using Fischer WLP 004 silicone-based thermal paste and secured in place with Kapton tape. The sensor data were also stored in the archive data storage.

The thermographic measurement system used to obtain comprehensive cavity surface temperature data consists of a FLIR T560 infrared camera, featuring a detector array of 640 $\times$ 480 pixel and equipped with a lens of 10 cm focal length. 
The corresponding FLIR Tools+ analysis software was used. This software allows for automated evaluation of maximum / minimum temperature spots in areas or predefined lines, which can are later used for further evaluations. The data can be exported as CSV files for further analysis.
\begin{figure}[t]
  \centering
  \includegraphics[width=0.47\textwidth]{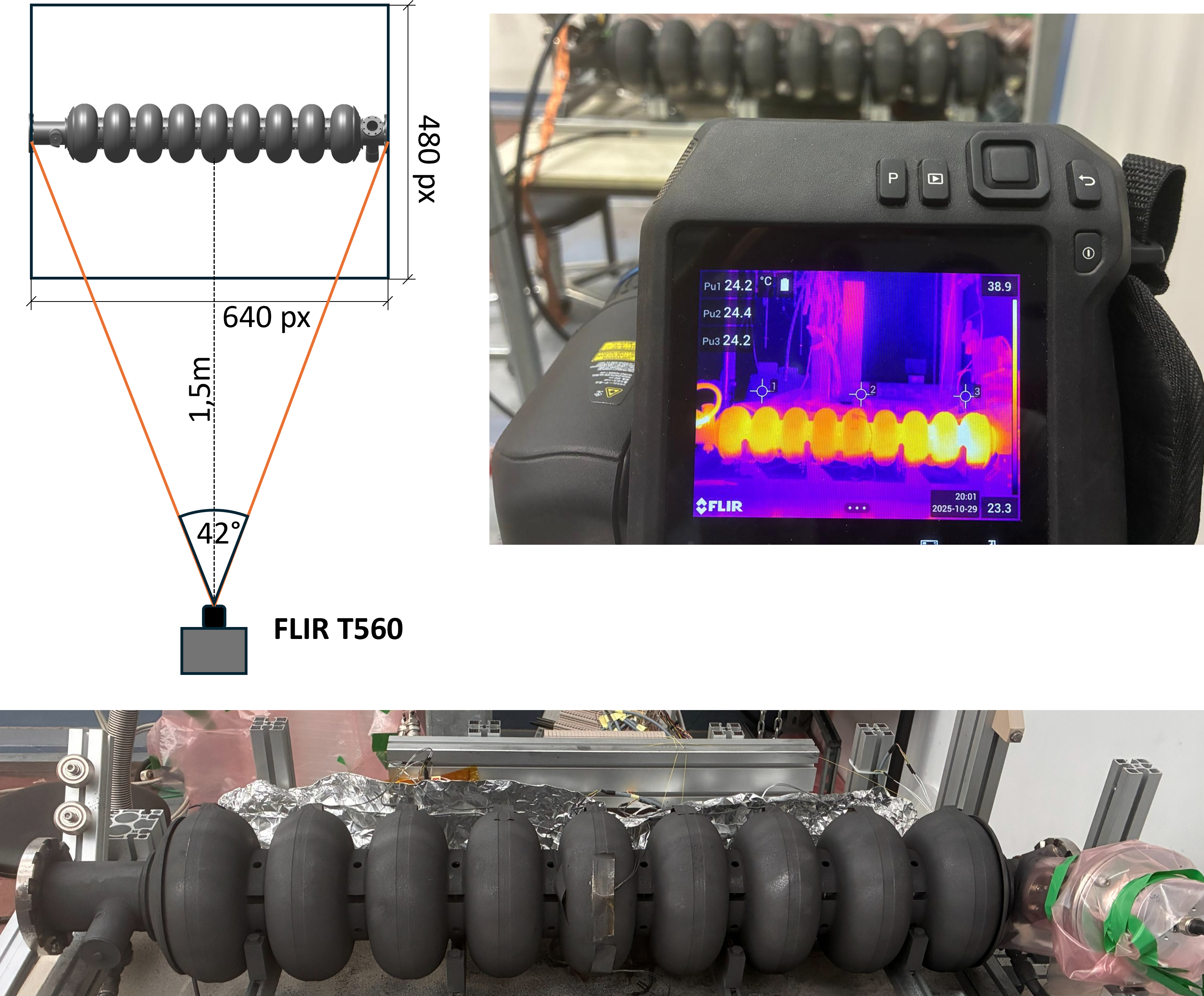} 
  \caption{Infrared thermography measurement setup. Top left: schematic representation of the setup; top right: over-the-shoulder view of the FLIR T560 infrared camera; bottom: TESLA 9-cell cavity with its surface fully coated with graphite.}
  \label{fig:2}
\end{figure}
The entire measurement setup is operated under atmospheric conditions and the cavity is exposed to air at both its inner and outer surfaces.

Figure~\ref{fig:2} shows the cavity in the measurement configuration. To obtain a spatially homogeneous surface with an emissivity coefficient close to 1, the cavity surface was coated with a commercial multi-purpose graphite spray ("Graphitspray", NIGRIN GmbH \& Co. KG, Osnabrück, Germany). To fit the entire cavity including the beam pipes within the camera's field of view, the camera was mounted on a rigid support at a distance of 1.5~m from the cavity (cf. Fig.~\ref{fig:2}). This results in a projected measurement area of approximately $(1.8 \times 1.8~\mathrm{mm}^2)$ per pixel (depending on the actually position on the cavity).
\begin{figure}[b]
  \centering
  \includegraphics[width=0.47\textwidth]{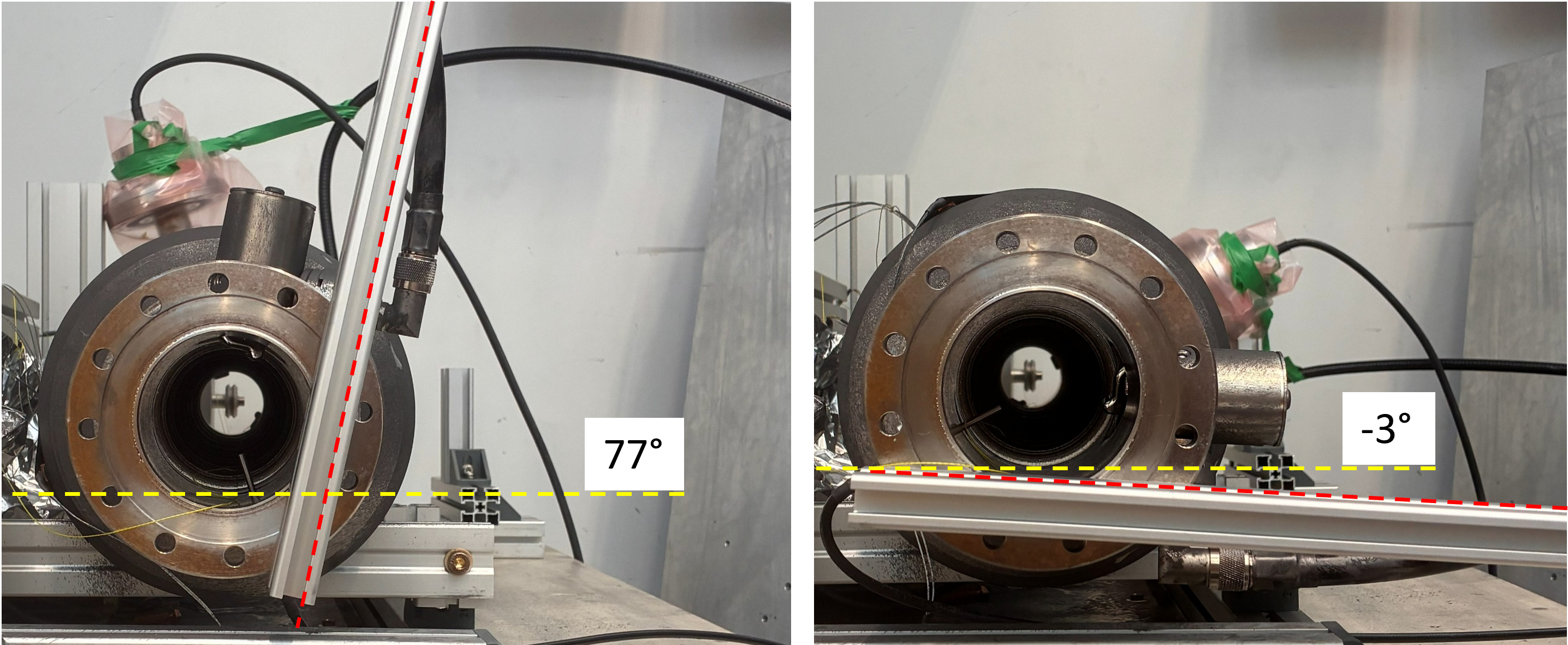} 
  \caption{80$^\circ$ rotatable setup with fixed mechanical stops for reproducible positioning. RF cables were mechanically supported to prevent connection issues during rotation. The camera views the setup from the right.}
  \label{fig:3}
\end{figure}
To obtain measurement data covering at least 50\% of the cavity surface, the cavity including its cabling was mounted on a support frame such that it was possible to rotate it by approximately ($80^\circ$). 

Figure~\ref{fig:3} shows the measurement arrangement at the two end positions defined by mechanical stops, viewed along the axis of rotation. 
A manual rotation of the cavity during the measurement between both end positions took approximately 5~s, which is a short span compared to the characteristic time scale of surface temperature changes of the cavity. 

\section{Calibration and Preliminary Analysis}
\begin{figure}[b]
  \centering
  \includegraphics[width=0.47\textwidth]{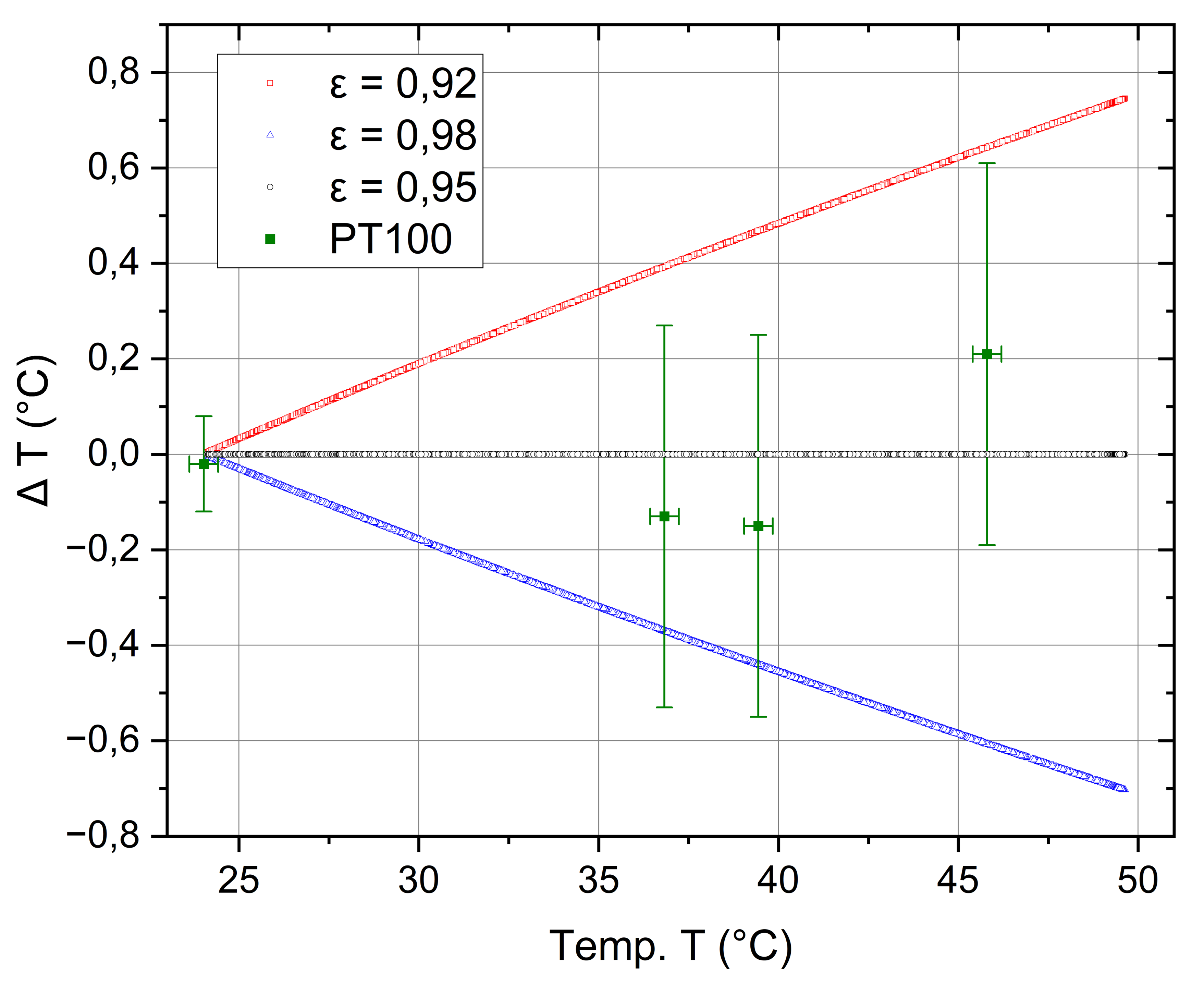} 
  \caption{Deviation of the infrared-measured temperature of a graphite-coated cavity surface region referenced to the PT100 readout in the region's center: $\Delta T = Tir - Tref_{11}$ for three different emissivity settings $\varepsilon=0.92, 0.95, 0.98$ (red, black, blue resp.) in the camera.
  The ambient temperature was set to $24^\circ\mathrm{C}$.
   All other camera settings were kept constant. The green data points represent the PT100 reference measurements, including the corresponding error bars.}
  \label{fig:4}
\end{figure}
To determine the emissivity of the cavity surface after application of the graphite coating, the IR setup has to be calibrated. 
Figure~\ref{fig:4} shows the variation in the evaluated surface temperature resulting from different emissivity settings in the FLIR software for a measurement of the coated surface up to $50^\circ$C. 
The data were evaluated using $\varepsilon = 0.92$, $0.95$, and $0.98$ (red, black, and blue, respectively). 
The remaining set parameters of the thermographic measurement were kept constant throughout the calibration and all subsequent measurements: a reflected temperature of $24^\circ\mathrm{C}$ (denoting the reflection of ambient black body radiation in overlay to own thermal radiation of a surface, which is a minor contribution for highly emissive surfaces), a camera-to-surface distance of $1.5~\mathrm{m}$, an atmospheric temperature of $24^\circ\mathrm{C}$, and a relative humidity of 60\%. 
Time-correlated PT100 readings from additional measurements, including their error bars, are shown in green. 
The data scatter around zero without an apparent systematic offset and are best consistent with an emissivity of $\varepsilon = 0.95$ within the measurement uncertainties.  
\begin{figure}[t]
  \centering
  \includegraphics[width=0.48\textwidth]{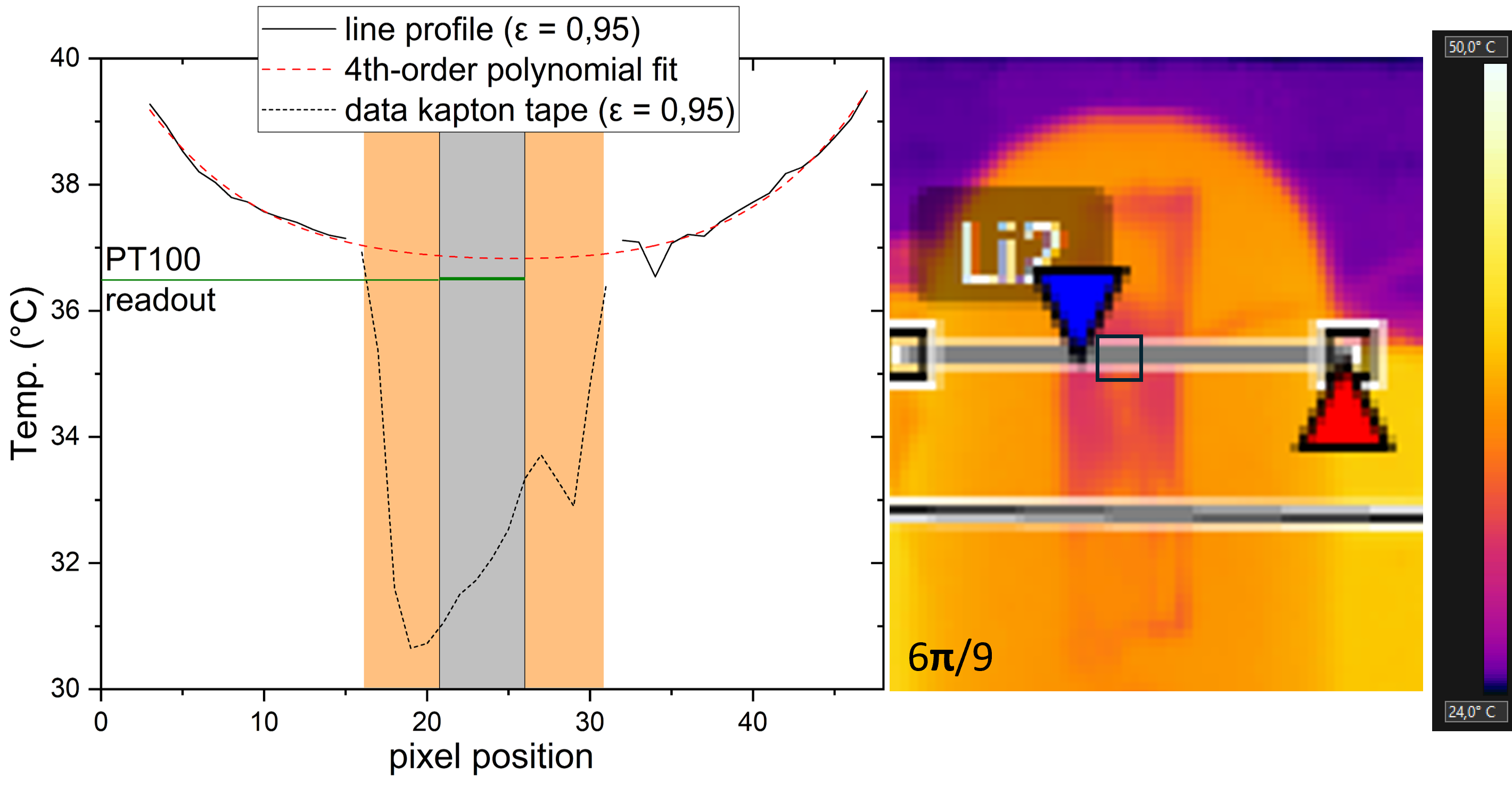} 
  \caption{Right: Thermal image of the middle cell during $6\pi/9$-mode heating, showing line profile Li2, the position of the PT100-equipped copper plate (black square), and minimum/maximum temperatures (blue/red triangles). Left: Temperature along Li2 at $\varepsilon=0.95$ (black), with fourth-order polynomial fits (red dashed) and PT100 reading (green). Excluded data are shown as black dashed lines, while Kapton tape and copper plate positions are highlighted in beige and gray, respectively.}
  \label{fig:5}
\end{figure}
Since the PT100 sensor mounted on the copper block is fixed with Kapton tape, this area has a different emissivity from the surrounding cavity surface. This is evident from the irregular temperature profile in the region covered by the Kapton tape, shown by the black dashed line for $\varepsilon = 0.95$. Therefore, for the correlation $\Delta T = T_\mathrm{IR} - T_{\mathrm{ref}11}$, the IR temperature ($T_\mathrm{IR}$) at the corresponding PT100 position ($T_{\mathrm{ref}11}$) has to be interpolated from measurements of the surrounding surface. Figure~\ref{fig:5} (right) shows the respective thermal camera image. The selected line profile data are exported from the FLIR software  for further analysis. The position of the copper block carrying the PT100 sensor is indicated(black square). Figure~\ref{fig:5} (left) shows the evaluated IR temperature as a function of pixel position along this line profile. The region corresponding to the Kapton tape, indicated by the beige bar, was excluded from the analysis. A fourth-order polynomial fit to the data (red dashed line) yielded a coefficient of determination ($R^2$) of 0.975, where values close to 1 indicate good agreement between the fit and the measured data. To validate the interpolation, the same fit was applied to a complete line profile of another cell without Kapton tape, measured at a comparable temperature and with a similarly symmetric iris-to-equator temperature distribution, yielding $R^2 = 0.988$. This supports the suitability of the fourth-order polynomial for the interpolation. 
\begin{figure}[t]
  \centering
  \includegraphics[width=0.38\textwidth]{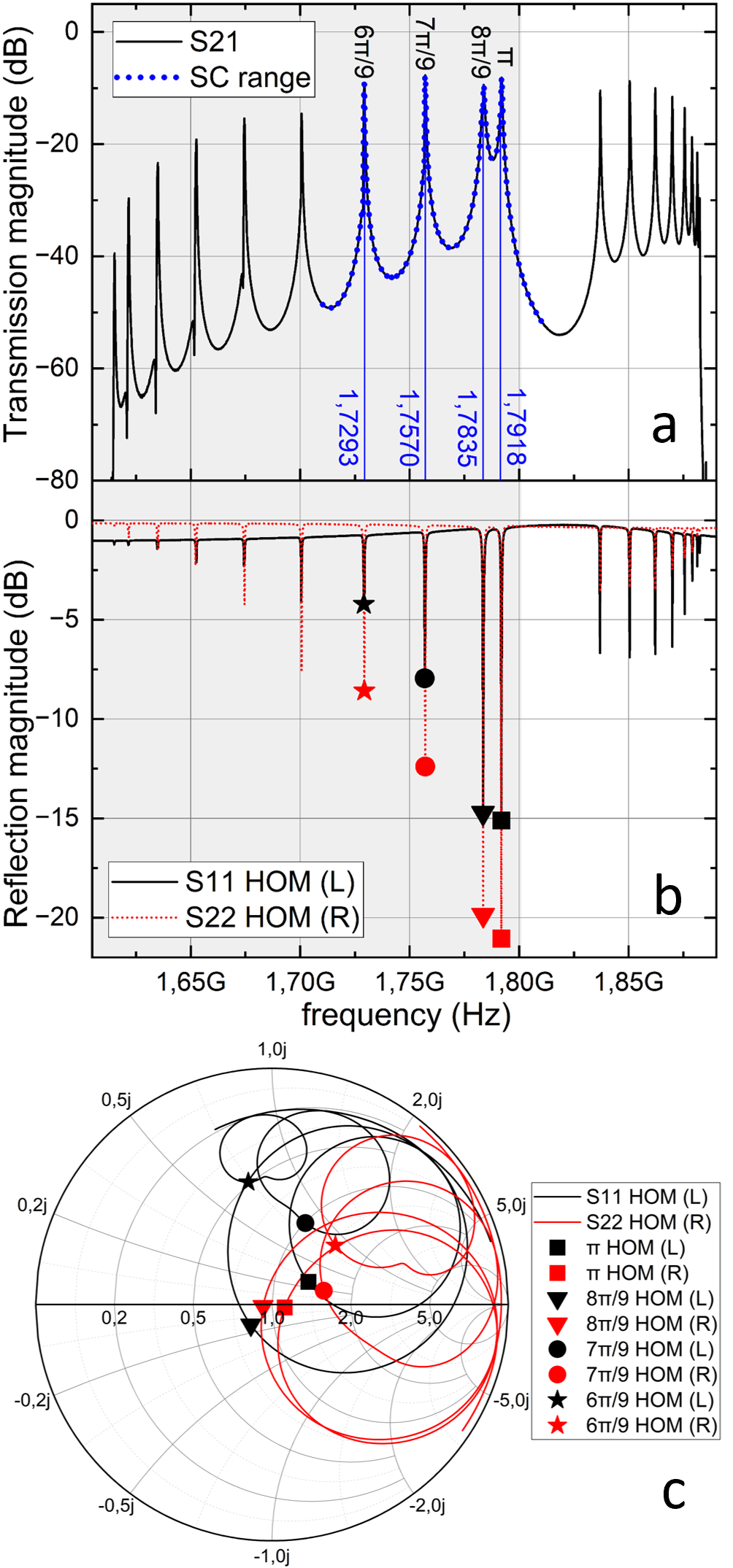} 
  \caption{VNA-measured scattering parameters of the first two dipole passbands of the TESLA 9-cell cavity. (a) Transmission magnitude $S_{21}$ over the frequency range from 1.60 to 1.88 GHz. The first dipole passband is shaded in gray, and the measured resonance frequencies of the $6\pi/9$, $7\pi/9$, $8\pi/9$, and $\pi$ modes are marked in blue. Blue dotted lines indicate the frequency range displayed in the Smith chart. (b) Reflection magnitudes $S_{11}$ (HOM (L), black) and $S_{22}$ (HOM (R), red dotted), with the resonance minima of the four investigated modes indicated. (c) Smith chart showing $S_{11}$ (black) and $S_{22}$ (red), with the positions of the four resonances marked.}
  \label{fig:6}
\end{figure}
Overall, the calibration indicates that an emissivity of $\varepsilon = 0.95$ provides good agreement with the reference measurements and is therefore used for the subsequent analysis. Since the following evaluation primarily focuses on relative temperature differences $\Delta T$, even deviations in the assumed surface emissivity of $\pm 0.03$ around $\varepsilon = 0.95$ have only a minor influence on the determined temperature differences.

Initially the frequency response of the cavity was characterised. Figure~\ref{fig:6} (a) shows the $S_{21}$ spectrum of the first two dipole passbands of the TESLA cavity, measured between the HOM couplers. The modes investigated in this work, the \(6\pi/9\), \(7\pi/9\), \(8\pi/9\), and \(\pi\), are labelled and their corresponding measured resonance frequencies are indicated in blue. As it can be inferred from Figures.~\ref{fig:6} a and b, the $\pi$, $8\pi/9$, $7\pi/9$ and $6\pi/9$ modes in the first dipole band are the most suitable ones due to their highest coupling. It is worth highlighting that the right HOM (R) coupler exhibits significantly stronger coupling in the first dipole passband than the left HOM (L) coupler, in contrast to the second dipole passband.
Figure~\ref{fig:6} (c) shows the Smith chart of $S_{11}$ for HOM (L) (black) and $S_{22}$ for HOM (R) (red) over the frequency range covering the four modes. The pronounced resonance circles are clearly visible, with the largest circles occurring for the $\pi$ and $8\pi/9$ modes, particularly for HOM (R). The resonance positions are also marked, corresponding to those in Fig.~\ref{fig:6} (b).
\begin{figure}[t]
  \centering
  \includegraphics[width=0.45\textwidth]{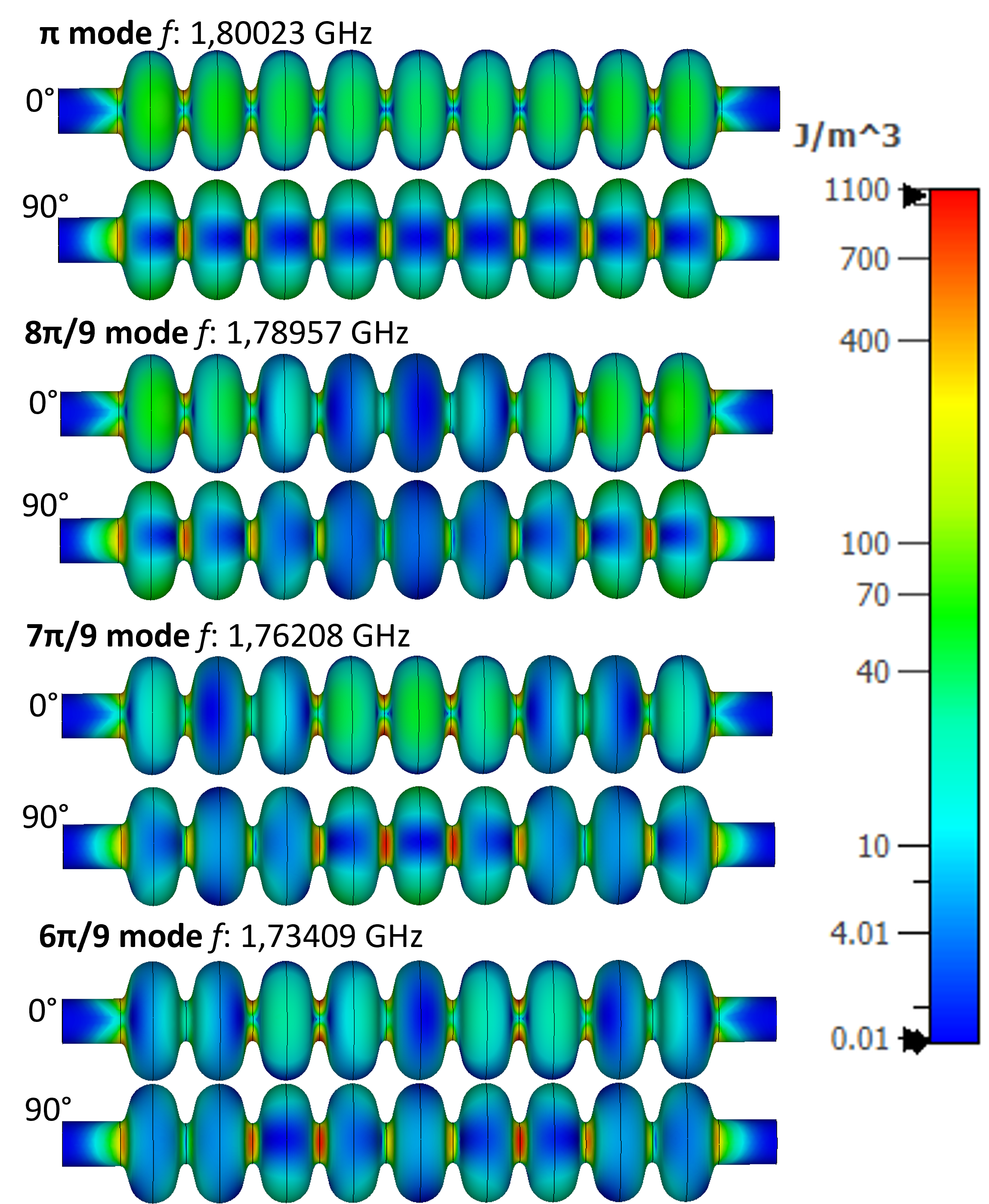} 
  \caption{CST simulation of a TESLA 9-cell cavity geometry based on B. Aune et al\cite{Aune2000_TESLA}. Shown are the surface magnetic energy density distributions of the \(6\pi/9\), \(7\pi/9\), \(8\pi/9\), and \(\pi\) modes of the first dipole passband and their calculate frequencies.}
  \label{fig:7}
\end{figure}
\begin{figure}[t]
  \centering
  \includegraphics[width=0.47\textwidth]{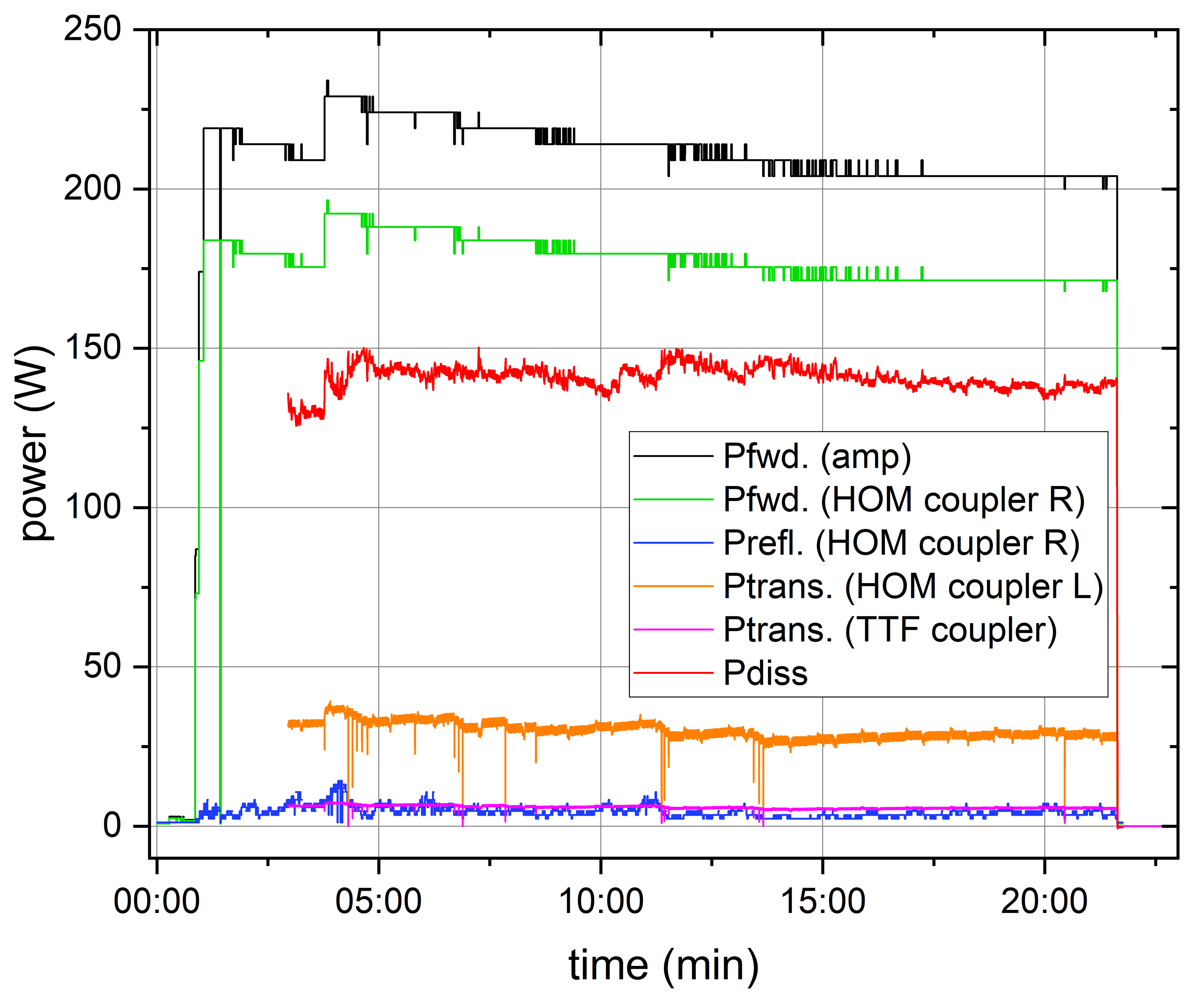} 
  \caption{Exemplary power balance for the 20-minute heating experiment in the $\pi$~mode. The pickup was not connected during this measurement. RF power meter data are missing for the first three minutes. Shown are $P_\mathrm{fwd}$ directly read from the amplifier (black), $P_\mathrm{fwd}$ at the right (R) HOM coupler, $P_\mathrm{refl}$ at the right (R) HOM coupler, transmitted power $P_\mathrm{trans}$ at the left (L) HOM coupler measured by the rf power meter (orange), transmitted power $P_\mathrm{trans}$ at the TTF coupler (pink), and the power dissipated in the cavity, $P_\mathrm{diss}$ (red).}
  \label{fig:8}
\end{figure}

Figure~\ref{fig:7} shows the surface magnetic energy density distributions of the $6\pi/9$, $7\pi/9$, $8\pi/9$, and $\pi$ modes of the first dipole passband, together with their corresponding calculated frequencies in GHz. For the simulations performed using CST, the TESLA cavity geometry based on B. Aune et al.~\cite{Aune2000_TESLA} was used, with an additional 10 cm of beam pipe on both sides. Each mode is shown in two sectional planes, denoted as $0^\circ$ and $90^\circ$. In addition to the expected dipole characteristics at the equator, a notable also expected feature is the enhancement of the magnetic energy density in the iris region. The maxima at the iris and equator are shifted azimuthally by $90^\circ$ relative to each other for each mode.

In order to illustrate power measurements figure~\ref{fig:8} shows an exemplary evaluation of the input and output powers during the 20-minute heating experiment in the $\pi$~mode. The forward and reflected power readings provided by the R\&S power amplifier were recorded. The amplifier power readout is limited to increments of $0.1$~dB, corresponding to power steps of approximately $5$~W at an output power of about $200$~W, as visible in the black $P_\mathrm{fwd,amp}$ curve. Accounting for the measured cable attenuation of $(0.76 \pm0.05)$~dB between the amplifier and the right (R) HOM coupler yields the forward power at the coupler, $P_\mathrm{fwd,HOM(R)}$, shown in green. Similarly, the reflected power measured at the amplifier was corrected for the same cable attenuation, resulting in $P_\mathrm{refl,HOM(R)}$, shown in blue. In addition, the output power at the left (L) HOM coupler, measured using an rf power meter and corrected for the corresponding cable attenuation, is shown in orange as $P_\mathrm{trans,HOM(L)}$. The transmitted power at the TTF coupler, $P_\mathrm{trans,TTF}$, was not measured directly, since a second rf power meter was not available. Instead, it was estimated from the ratio of $S_{21}$ and $S_{41}$ obtained from the measured S-parameter spectrum. The dissipated power $P_{\mathrm{diss}}$ (red) is thus determined from energy conservation as $P_{\mathrm{diss}} = P_{\mathrm{fwd,HOM(R)}} - P_{\mathrm{ref,HOM(R)}} - P_{\mathrm{trans,HOM(L)}} - P_{\mathrm{trans,TTF}}$.
\section{Results}
Since the measurements were performed in air and with a modified surface emissivity, the analysis does not focus on absolute changes in the cavity surface temperature, as the heat transfer conditions, particularly convection and thermal radiation, are not directly comparable to those of a cavity installed in a cryomodule. Instead, the first subsection investigates the azimuthal temperature differences at the cell equators to characterize the dipole-like field distribution and its influence on the temperature distribution. The second subsection examines temperature differences along the cavity contour, and the third subsection examines the azimuthal temperature distribution in the iris region.
\subsection{Azimuthal Equator Analysis}
This section investigates the extent to which the dipole mode characteristics influence the azimuthal temperature distribution along the cell equator.

\begin{figure}[b]
  \centering
  \includegraphics[width=0.48\textwidth]{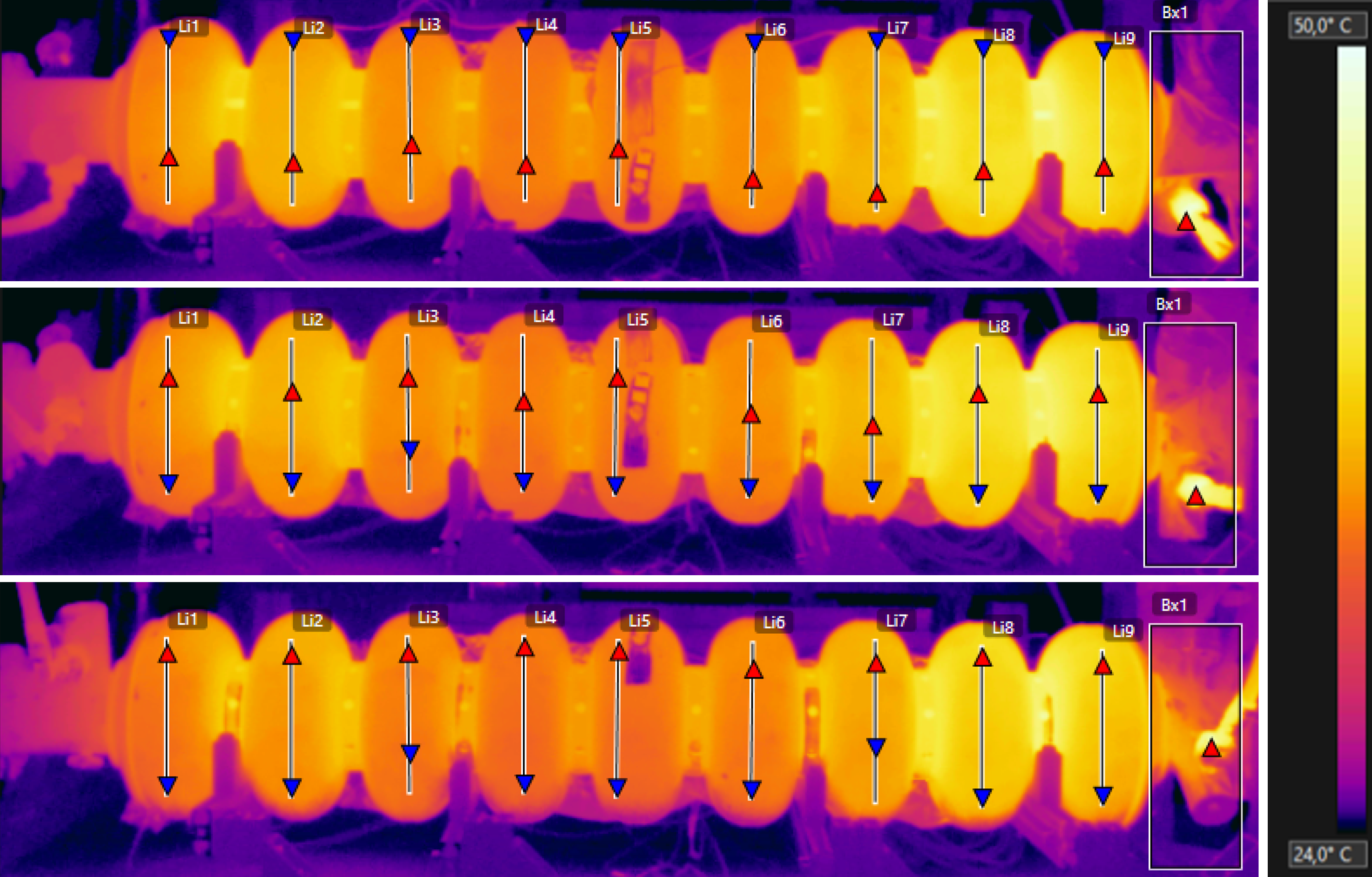} 
  \caption{Infrared images taken during the final rotation of the 20 min heating experiment in the $\pi$~mode. Top: initial position at the lower rotation limit, corresponding to the orientation shown in Fig.~\ref{fig:3} (right). Middle: intermediate position during rotation. Bottom: final position at the opposite rotation limit, corresponding to the orientation shown in Fig.~\ref{fig:3} (left). Nine line profiles (Li\#) are shown at the equator position of each cell, with the profile for cell 5 slightly shifted from the equator.}
  \label{fig:9}
\end{figure}

For this purpose, Fig.~\ref{fig:9} shows an infrared thermal image of the TESLA cavity at the end of a 20-minute heating experiment in the $\pi$~mode.
The upper image was taken immediately before the final rotation, after 20 minutes of heating, with the cavity in the orientation shown in Fig.~\ref{fig:3} (right). The middle image was recorded during the rotation of the cavity, while the lower image was taken immediately after the rotation, corresponding to the orientation shown in Fig.~\ref{fig:3} (left). The line profiles Li1-Li9 were placed approximately along the equator of the respective cells (Li5 has a slightly offset due to the sensors and kapton tape) and can be extracted from the thermal images for quantitative analysis. The red and blue triangles indicate the hottest and coolest measurement regions, respectively, along each line profile. During the rotation, the positions of the hot spots were observed to move with the cavity while remaining approximately within the same azimuthal plane. This behavior is in good agreement with the simulated surface magnetic field distribution of the $\pi$~mode shown in Fig.~\ref{fig:9}, which exhibits maxima at the same azimuthal angle in all cells. 
The input power during this heating experiment can be obtained from Fig.~\ref{fig:8}, with an average input power of ($177 \pm 5$)~W applied to the right (R) HOM port, resulting in a dissipated power of $P_\mathrm{diss} = (140.0 \pm 7.5)$~W. An uncertainty of $\pm 7.5$~W was assigned to the derived power values. This estimate includes the specified amplifier power uncertainty of approximately $\pm 2.5$~W, uncertainty in the damping values and an additional conservative contribution accounting for systematic uncertainties. These include, in particular, possible changes in cable attenuation due to temperature variations as well as frequency shifts during cavity heating, since the rf cables were calibrated at room temperature ($24^\circ$~C). 
\begin{figure}[t]
  \centering
  \includegraphics[width=0.45\textwidth]{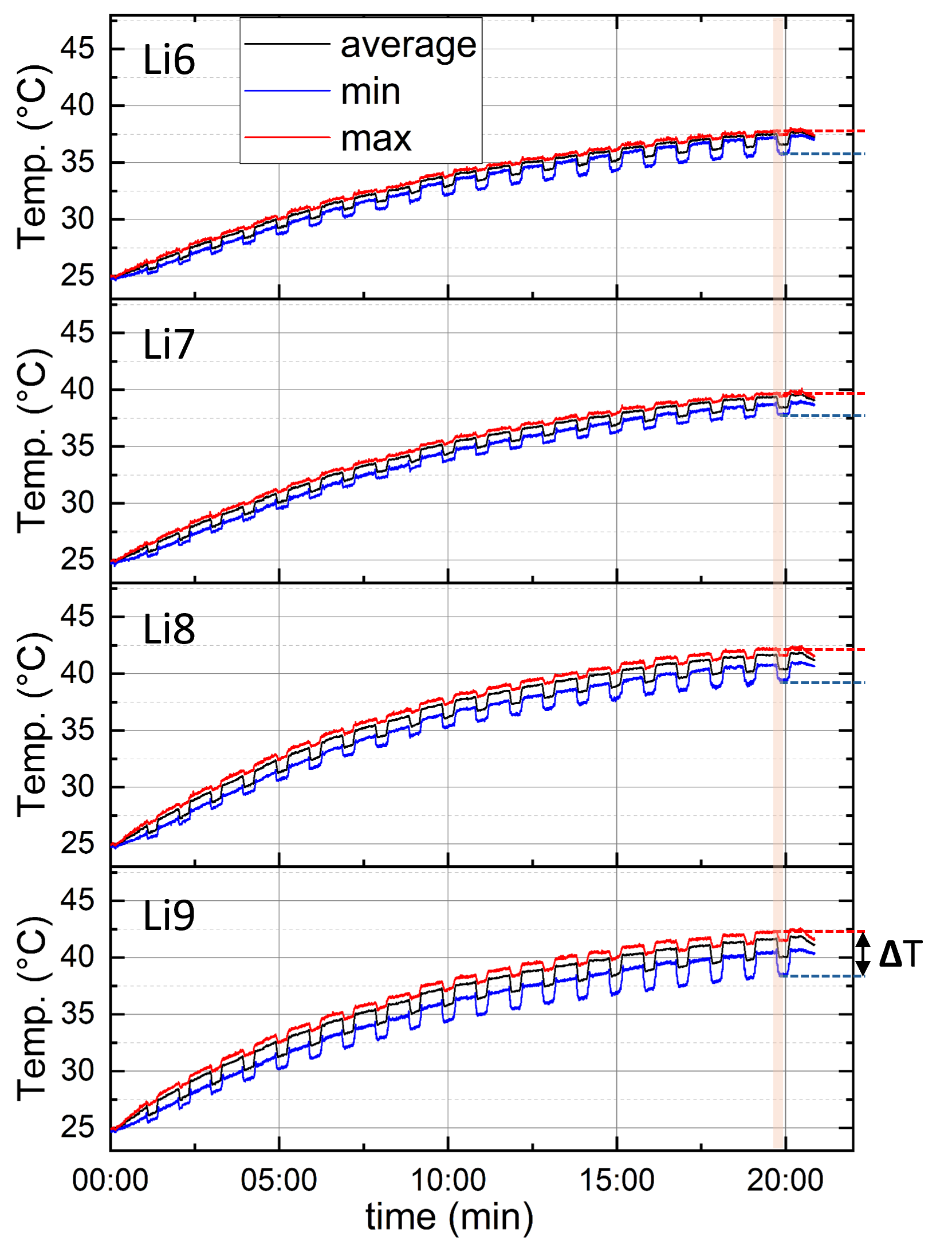} 
  \caption{Temperature evolution of line profiles Li6–Li9 during $\pi$-mode heating. Maximum, average, and minimum temperatures are shown in red, black, and blue, respectively. The beige region marks the final rotation shown in Fig.~\ref{fig:11}. Dashed red and blue lines indicate $\Delta T$ between the maximum before and minimum after rotation.}
  \label{fig:10}
\end{figure}
\begin{table}[b]
    \centering
    \caption{Temperature differences $\Delta T$ along the line profiles Li6--Li9 at different heating times using the $\pi$~mode.}
    \label{tab:1}
    \begin{tabular}{l|cccc}
        & \multicolumn{4}{c}{$\Delta T$ (K)}\\
        \midrule
        Line profile & 5 min & 10 min & 15 min & 20 min \\
        \midrule
        Li6 & 1.6 & 1.9 & 2.0 & 2.0 \\
        Li7 & 1.6 & 1.9 & 1.9 & 1.9 \\
        Li8 & 2.2 & 2.6 & 2.7 & 2.8 \\
        Li9 & 2.9 & 3.3 & 4.0 & 4.0 \\
        \bottomrule
    \end{tabular}
\end{table}
Figure~\ref{fig:10} shows the maximum (red), minimum (blue), and average (black) temperatures of line profiles Li6--Li9 as a function of heating time. Among these, Li9 exhibits both the highest absolute temperature of $42.5^\circ$~C and the largest azimuthal temperature difference along the cell equator, reaching $\Delta T = 4$ K. The temperature difference $\Delta T$ is determined from the maximum temperature immediately before rotation and the minimum temperature immediately after rotation, assuming sufficient thermal inertia to neglect temperature changes during the approximately 5 s rotation period. An example of this determination of $\Delta T$ is illustrated in the diagram by the red and blue dashed lines during the rotation interval highlighted in beige.

The evolution of the temperature differences for these four line profiles at 5-minute intervals is summarized in Table~\ref{tab:1}. The temperature spread initially increases rapidly and reaches an approximately steady value after 15 min. Despite a further rise in absolute temperature, the spread remains nearly constant. Furthermore, cells with lower absolute temperature increases exhibit smaller temperature spreads. This is evident from line profiles Li6 and Li7, which reach max temperatures of 38 °C and 40 °C, respectively, but show an azimuthal temperature spread of only about 2~K along the cell equators.
\begin{figure}[b]
  \centering
  \includegraphics[width=0.48\textwidth]{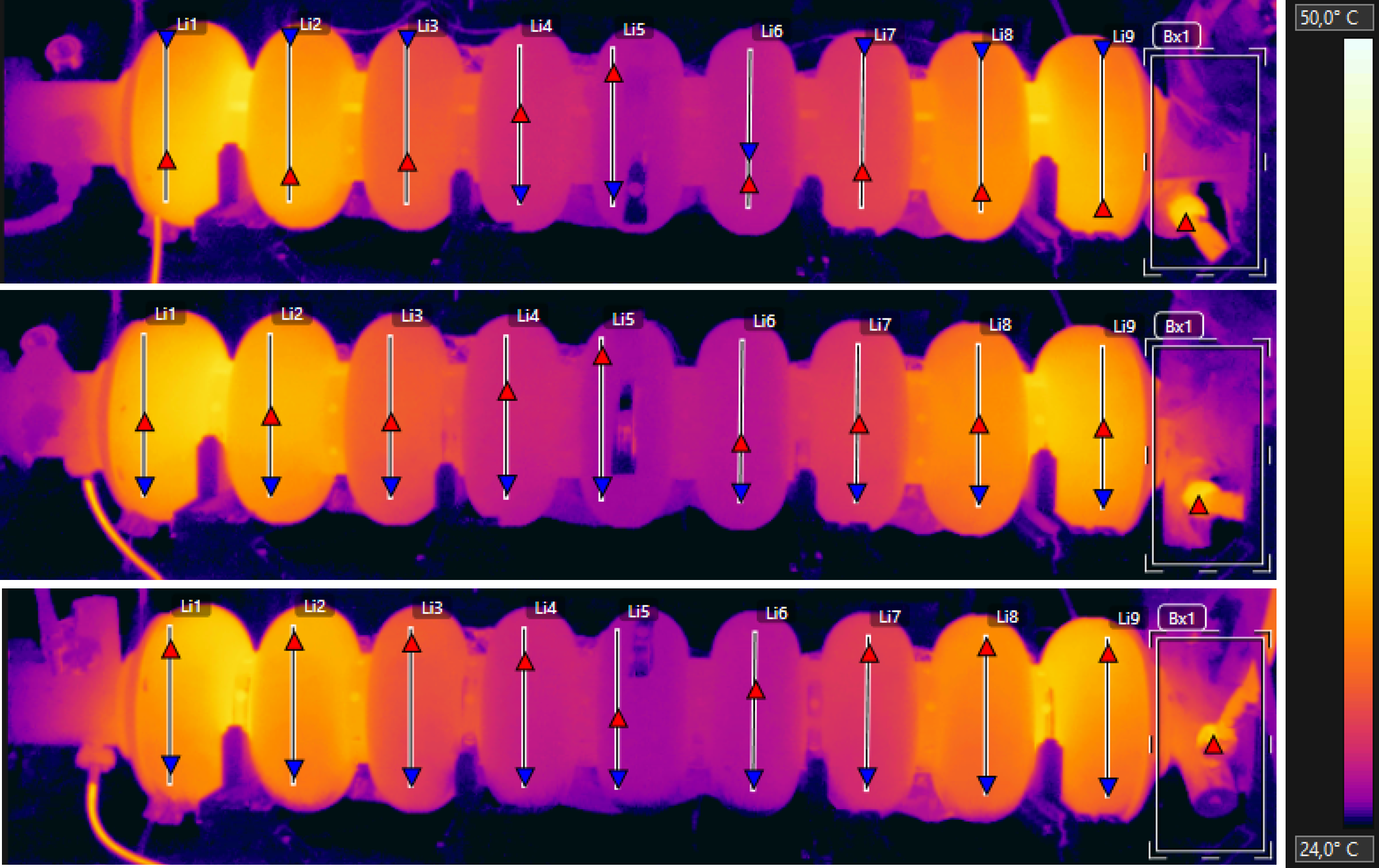} 
  \caption{IR images taken after 10~min rf heating in the $8\pi/9$~mode. Top: initial position at the lower rotation limit,  Middle: intermediate position during rotation. Bottom: final position at the opposite rotation limit. Nine line profiles (Li\#) are shown at the equator position of each cell.}
  \label{fig:11}
\end{figure}

To contrast the more homogeneous cell-to-cell surface magnetic energy density distribution of the $\pi$~mode, the $8\pi/9$~mode is also investigated using the same evaluation method.
Figure \ref{fig:11} shows thermal images after 10 minutes of heating in the $8\pi/9$~mode  before, during, and after rotation, with the line profiles positioned as previously described. The temperature color scale reveals significantly stronger heating of the end cells compared to the inner cells. This agrees well with the simulated surface magnetic energy density distribution, which also exhibits higher values in the end cells of the TESLA cavity geometry (see Fig.~\ref{fig:7}).
During this experiment, the average rf power coupled into the cavity via HOM (R) was 190~W, with a low reflected power of 2~W, while an average of 7~W was extracted through HOM (L). In addition to the 7~W extracted through the TTF coupler, the power transmitted through the pickup must also be considered, as indicated by the heating of its thin rf cable visible in the thermal image. Based on the measured S-parameter spectrum, the average power extracted through the pickup is estimated to be 9~W, resulting in a dissipated rf power in the cavity of $P_{\mathrm{diss}} = (165.0 \pm 7.5)~\mathrm{W}$.
\begin{figure}[t]
  \centering
  \includegraphics[width=0.45\textwidth]{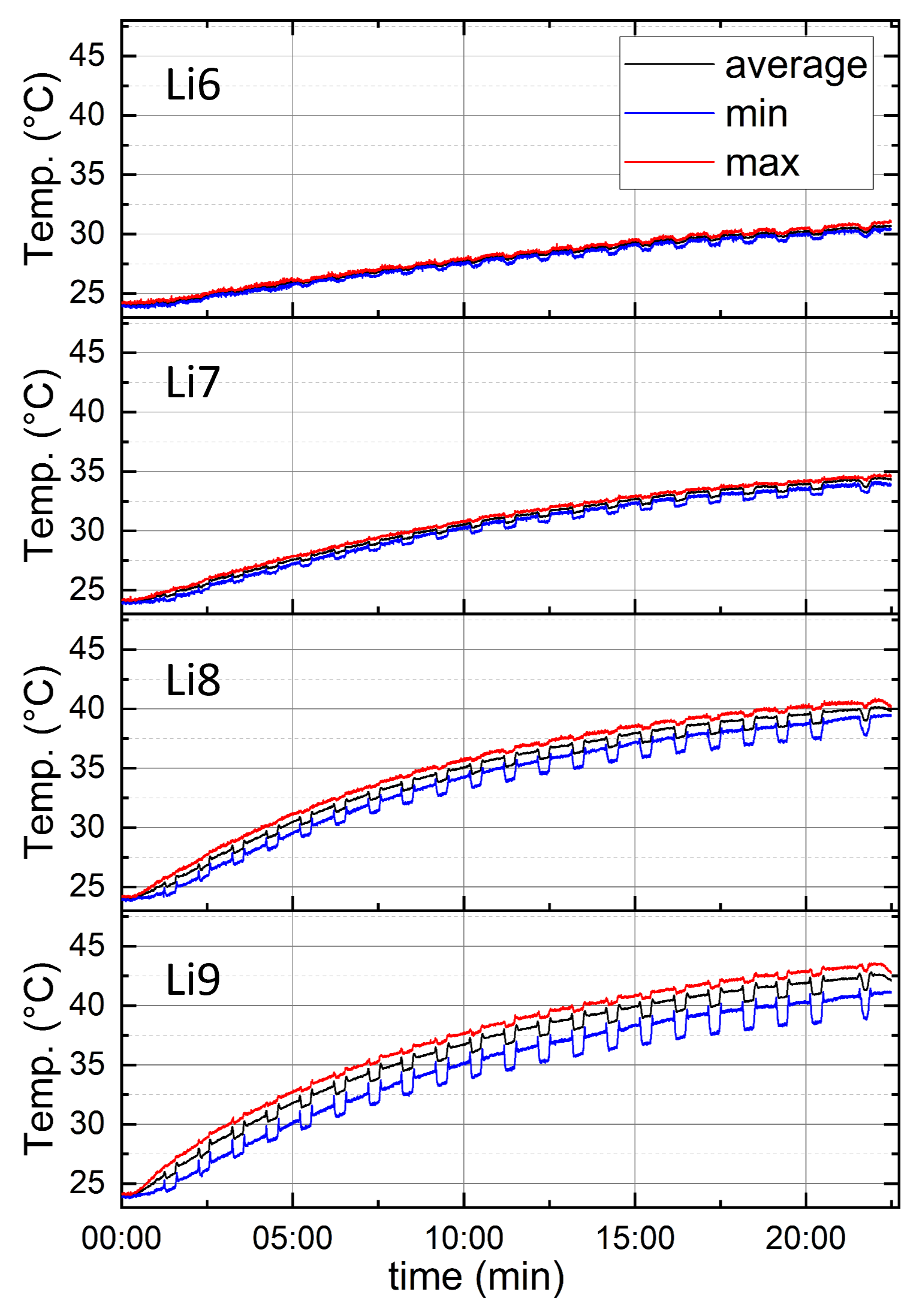} 
  \caption{Temperature evolution of line profiles Li6–Li9 during $8\pi/9$-mode heating. Maximum, average, and minimum temperatures are shown in red, black, and blue, respectively.}
  \label{fig:12}
\end{figure}

The pronounced differences in surface magnetic energy density between the cells are clearly reflected in the varying maximum equator temperatures, as well as in larger azimuthal temperature differences $\Delta T$ along the cell equators. This behavior is illustrated in Fig.~\ref{fig:12}. After 20 min of heating, Li9 reaches a temperature of 43 °C, approximately the same as in the $\pi$-mode experiment. The maximum temperatures of Li8, Li7, and Li6 decrease progressively, reaching 40.5 °C, 34.5 °C, and 31 °C, respectively. The temperature spread develops rapidly after rf power is applied to the cavity and increases only slightly over the remainder of the measurement. Table~\ref{tab:2} summarizes the temperature differences $\Delta T$ for Li6--Li9 at 5-minute intervals. 
\begin{table}[t]
    \centering
    \caption{Temperature differences $\Delta T$ along the line profiles Li6--Li9 at different heating times using the $8\pi/9$~mode.}
    \label{tab:2}
    \begin{tabular}{l|cccc}
        & \multicolumn{4}{c}{$\Delta T$ (K)}\\
        \midrule
        Line profile & 5 min & 10 min & 15 min & 20 min \\
        \midrule
        Li6 & 0.8 & 0.8 & 0.9 & 1 \\
        Li7 & 0.9 & 1.1 & 1.2 & 1.3 \\
        Li8 & 2.2 & 2.6 & 2.7 & 2.8 \\
        Li9 & 3.6 & 4.1 & 4.3 & 4.5 \\
        \bottomrule
    \end{tabular}
\end{table}
The largest azimuthal temperature difference $\Delta T$ is observed along the equator of Li9, reaching 4.5~K after 20~min. After 10~min, the temperature spread had already reached 4.1~K, increasing by only 0.4~K over the remaining 10~min.
\subsection{Temperature distribution along the cavity contour}
\begin{figure}[b]
  \centering
  \includegraphics[width=0.46\textwidth]{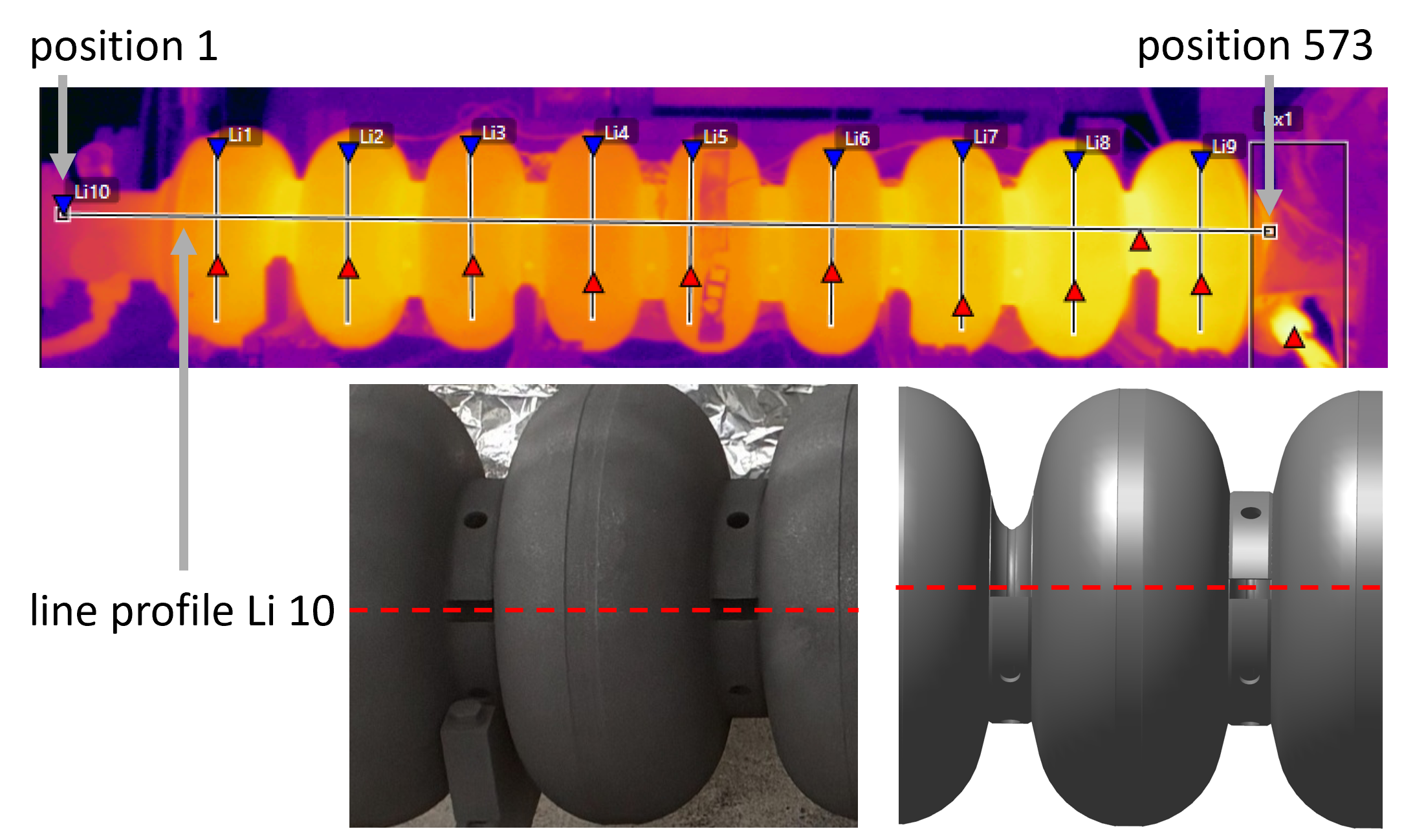} 
  \caption{Position of the horizontal IR line profile Li10 across the cavity (upper panel) and detailed view of the gap between the stiffening rings with the corresponding CAD model of the iris region (lower panel). The red dashed line indicates the position of Li10.}
  \label{fig:13}
\end{figure}
Since, as expected, the simulations predict maxima in the surface magnetic energy density near the cell iris regions, and an initial inspection of the IR images revealed enhanced heating in these regions, this section focuses on the temperature distribution along the TESLA cavity contour. Although the welded stiffening rings partially obstruct the view of the iris regions, the gaps between them are sufficiently wide to allow direct observation of the cavity surface with the IR camera. The upper panel of Fig.~\ref{fig:13} shows the position of line profile Li10, which extends horizontally across the entire cavity, passing directly over the gaps between the stiffening rings when the cavity is at its lowest rotational position. The positions 1 and 573 of the line profile are marked for reference in the subsequent analysis. The lower panel shows a photograph of the cavity with a view of the gap, alongside a CAD model of the iris region in which one half-shell is hidden to better illustrate the cavity contour. The position of Li10 is indicated by the red dashed line. 

\begin{figure}[t]
  \centering
  \includegraphics[width=0.48\textwidth]{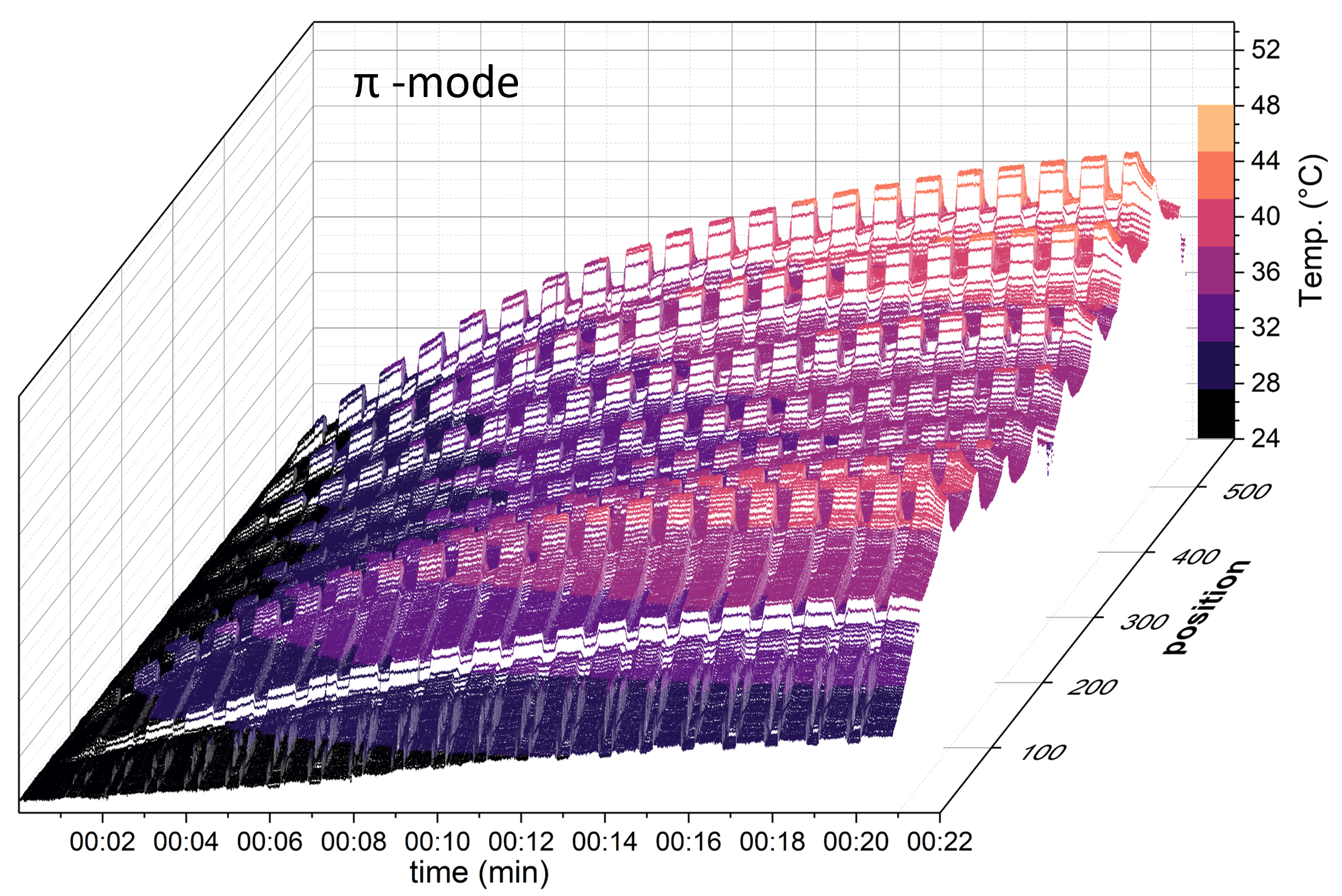} 
  \caption{3D representation of the 20 minute rf heating process using the $\pi$~mode, showing the temperature evolution along Li10 as a function of time and pixel position.}
  \label{fig:14}
\end{figure}

The 3D plot in Fig.~\ref{fig:14} shows the complete rf heating process using the $\pi$~mode along Li10, with temperature plotted as a function of time and pixel position.
Pronounced temperature maxima are observed, which, as shown below, correlate with the iris positions between adjacent cells. The largest abrupt temperature changes occur during the periodic cavity rotations, when the gaps between the stiffening rings move out of the line profile. The resulting lower temperatures correspond to the surfaces of the stiffening rings themselves.

\begin{figure}[b]
  \centering
  \includegraphics[width=0.46\textwidth]{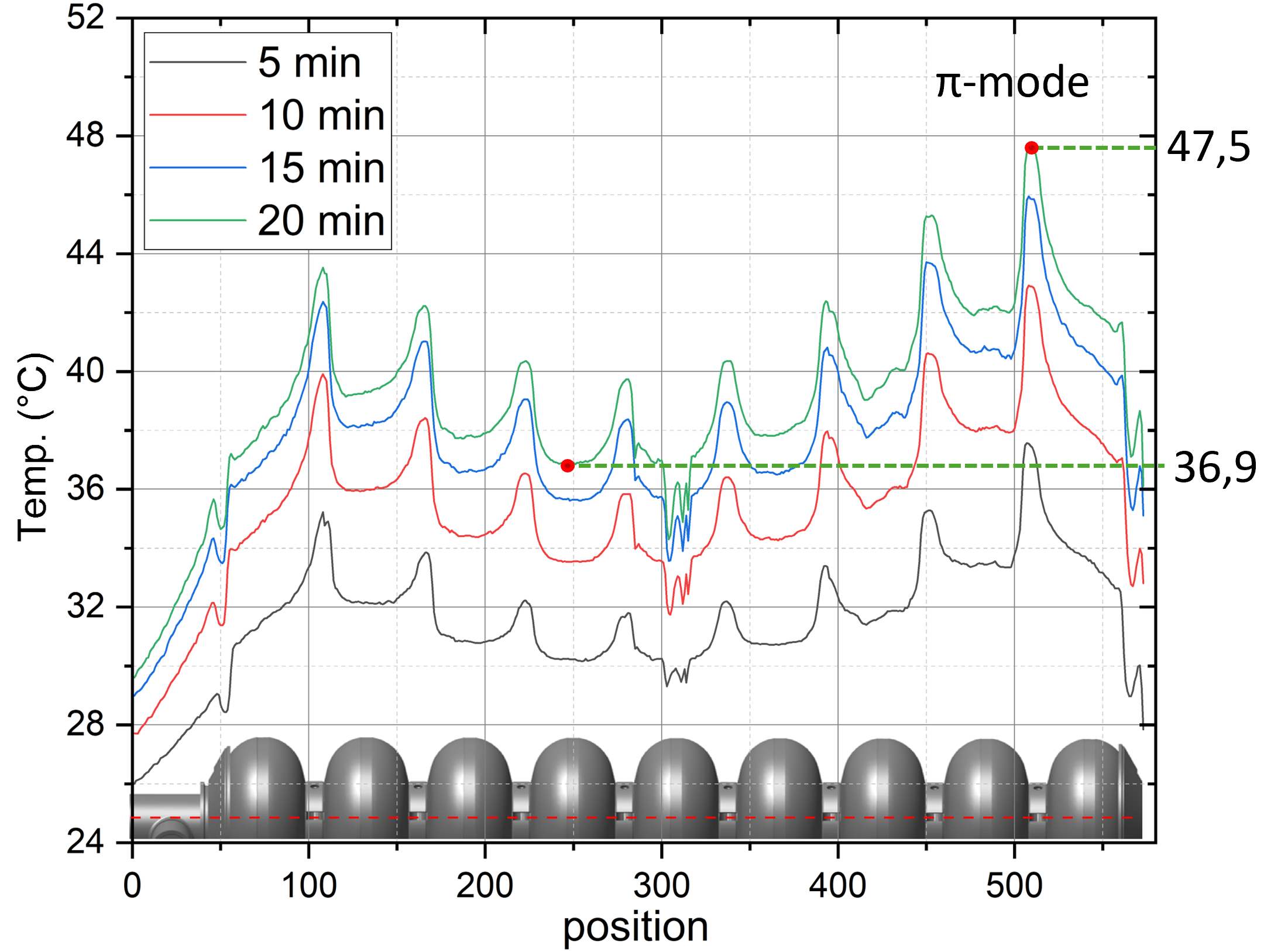} 
  \caption{Temperature profiles along Li10 after 5 min (black), 10 min (red), 15 min (blue), and 20 min (green) of rf heating using the $\pi$~mode. The aligned CAD model below illustrates the corresponding cavity geometry and iris positions.}
  \label{fig:15}
\end{figure}
Figure~\ref{fig:15} shows four temperature profiles recorded at 5-minute intervals, resolving the temperature distribution along the cavity contour, including the iris regions.
An aligned CAD model is included below the plot fore comparison with the cavity geometry. Pronounced temperature maxima at the iris positions are already visible after 5 min (black), with subsequent profiles shown after 10 min (red), 15 min (blue), and 20 min (green). After 20 min, the temperature spread along the entire cavity contour is determined from the marked extrema, with a maximum of 47.5~$^\circ$~C and a minimum of 36.9~$^\circ$~C, corresponding to $\Delta T = 10.6$ K. The corresponding temperature differences of $\Delta T = 7.4$ K, 9.4 K, and 10.3 K after 5, 10, and 15 min, respectively, indicate that the temperature spread gradually approaches a steady value. 

To estimate the overall temperature spread across the entire cavity after 20 min of heating, the minimum temperature of 36.9 °C was correlated with a measurement from Li4 at approximately the same location (see Figs.~\ref{fig:13} and \ref{fig:9}). The additional azimuthal temperature difference of $\Delta T = 2$ K determined from Li4 must therefore be taken into account, resulting in an overall temperature spread of $\Delta T = 12.6$ K. 

In general, the middle cells exhibit lower temperatures than the outer cells, and this effect is even more pronounced in the $8\pi/9$~mode. Therefore, mode alternation can be considered as a means of counteracting this temperature distribution, similar to the approach described in Ref.~\cite{wk7y-8n6d}. With a view to future investigations, the results for the $7\pi/9$ and $6\pi/9$~modes are presented in the following, as these modes appear promising for selectively heating the middle cells while also exhibiting sufficient coupling.

\begin{figure}[b]
  \centering
  \includegraphics[width=0.40\textwidth]{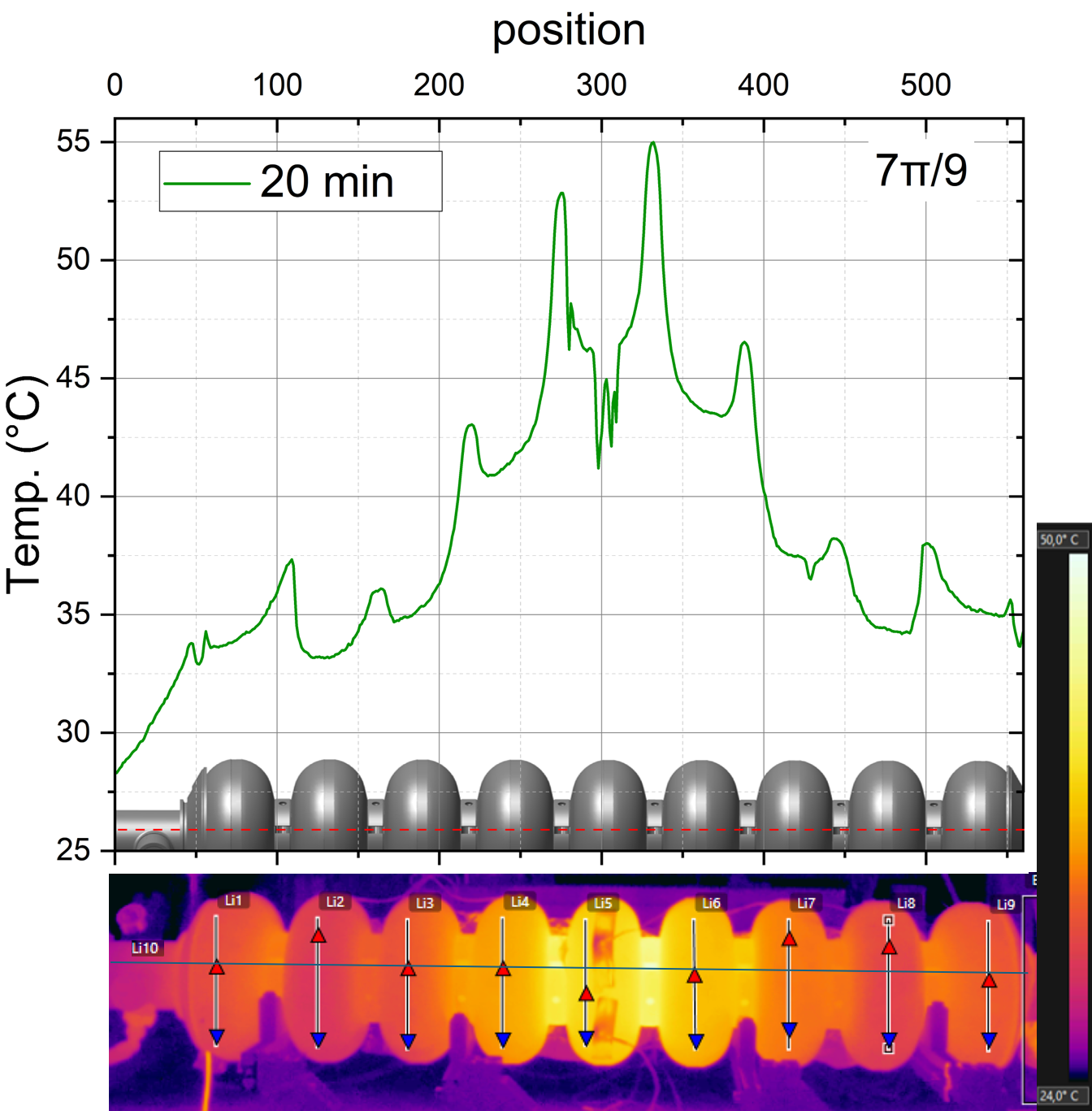} 
  \caption{Top: Temperature profile as a function of position after 20 minutes of rf heating. Bottom: Thermal image of the heating experiment with the indicated line profile Li10 position during rf heating using the $7\pi/9$~mode.}
  \label{fig:16}
\end{figure}
\begin{figure}[t]
  \centering
  \includegraphics[width=0.40\textwidth]{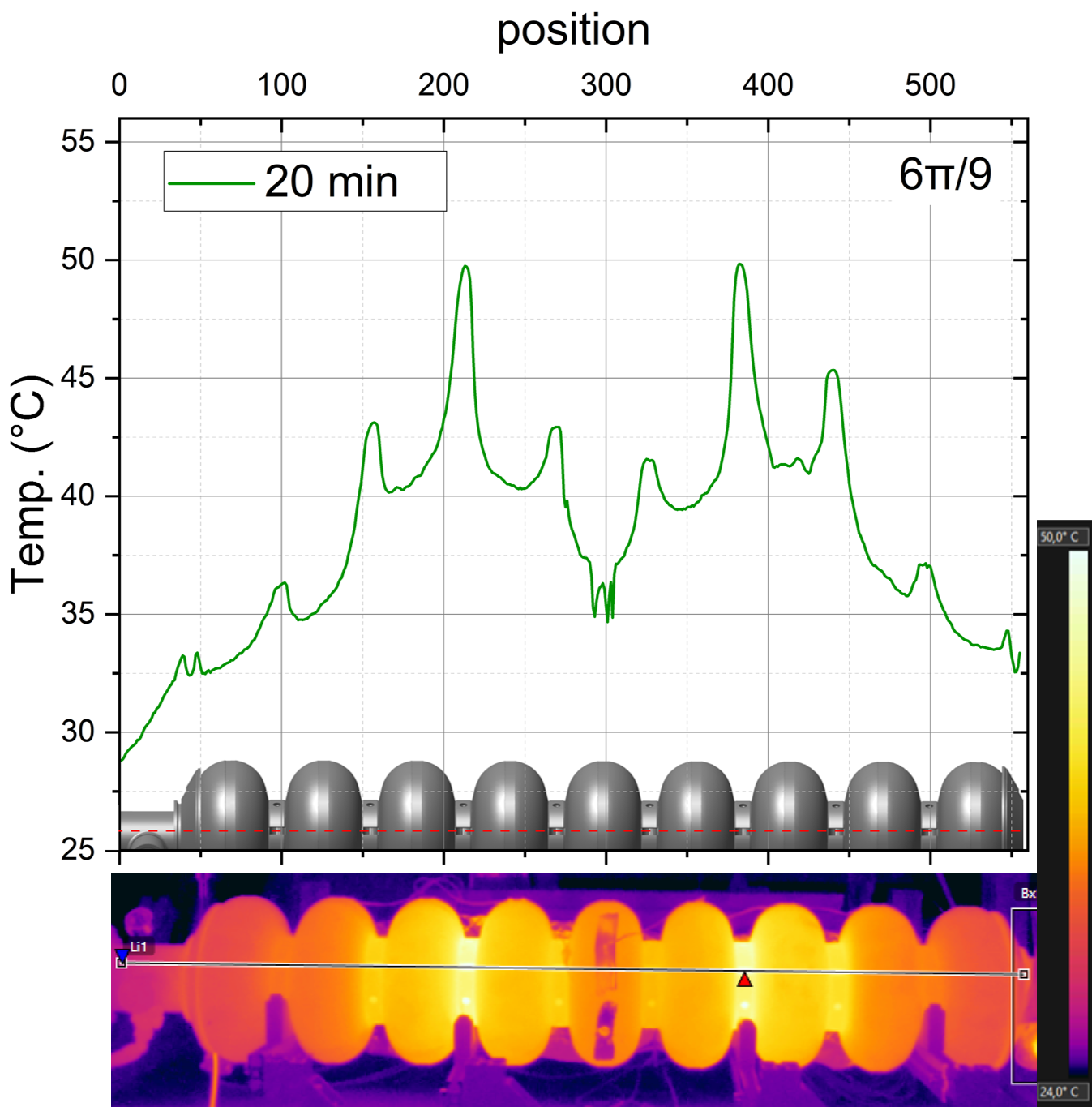} 
  \caption{Top: Temperature profile as a function of position after 20 minutes of rf heating. Bottom: Thermal image of the heating experiment with the line profile Li1 during rf heating using the $6\pi/9$~mode.}
  \label{fig:17}
\end{figure}
Figures~\ref{fig:16} and \ref{fig:17} show the thermal images in the lower panels, with the corresponding line profiles extending across the gap between the stiffening rings. The upper panels show the temperature profiles as a function of position. The average dissipated rf powers were $P_{\mathrm{diss}} = (154.0 \pm 7.5)~\mathrm{W}$ for the $7\pi/9$~mode and $P_{\mathrm{diss}} = (127.0 \pm 7.5)~\mathrm{W}$ for the $6\pi/9$~mode, with corresponding forward powers of $P_{\mathrm{fwd,HOM(R)}} = (201.5 \pm 7.5)~\mathrm{W}$  and ($193.0 \pm 7.5)~\mathrm{W}$, respectively, at the HOM coupler port (R).

It should be noted that the rf power dissipation is strongly concentrated in individual iris regions for those modes, as expected from the CST simulations shown in Fig.~\ref{fig:9}. These regions exhibit not only the highest absolute temperatures but also the largest temperature gradients within a cell. The temperature differences relative to the corresponding equator temperatures of the same cells are significantly larger than those observed for the $\pi$~mode.

\subsection{Azimuthal Iris Analysis}
\begin{figure}[t]
  \centering
  \includegraphics[width=0.46\textwidth]{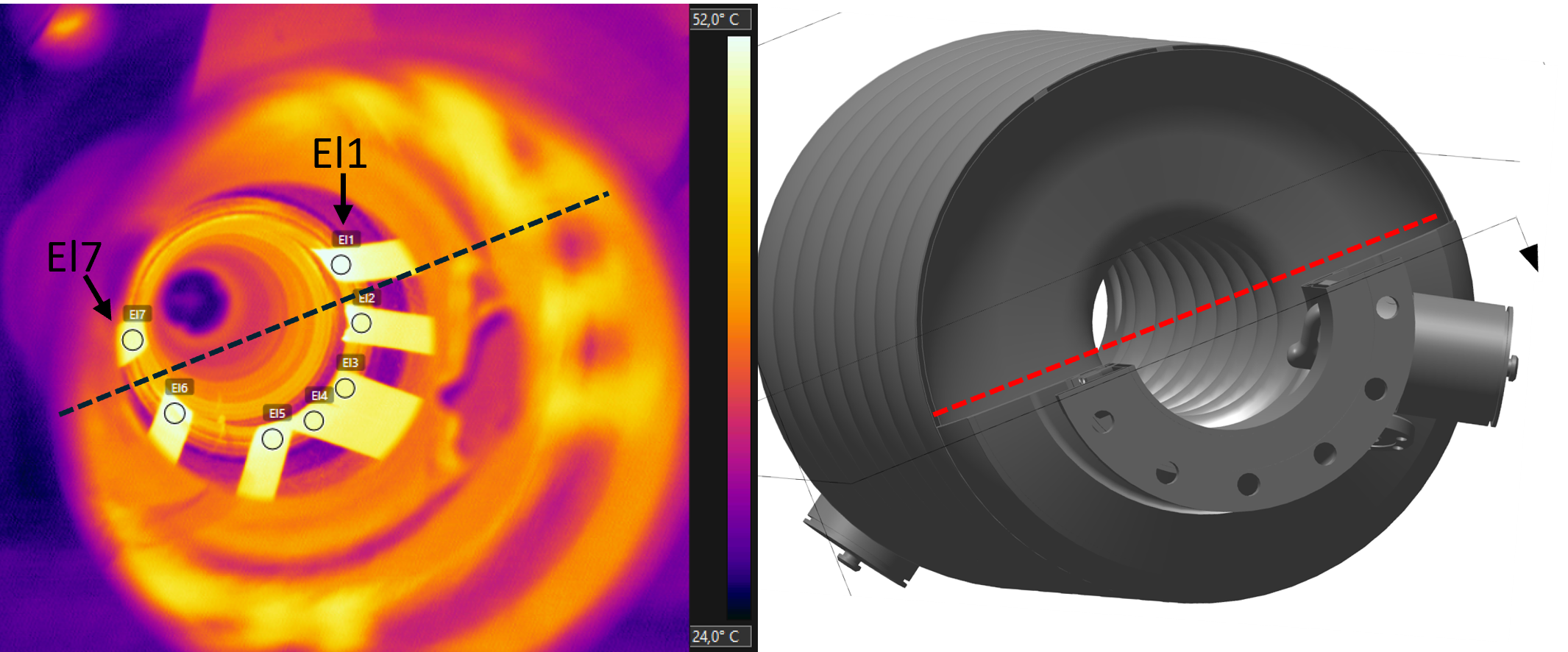} 
  \caption{IR image of the iris region on the pickup side during rf heating using the $8\pi/9$~mode (left), showing seven measurement positions (El1--El7) on PTFE adhesive tape applied to the niobium surface. The black dashed line indicates the azimuthal plane of the slots in the external stiffening rings. A partially sectioned CAD model of the first cell (right) illustrates the corresponding iris geometry, with the section plane highlighted in red.}
  \label{fig:18}
\end{figure}

According to the CST simulations, the gradient of the surface magnetic energy density is most pronounced in the iris regions (see Fig.~\ref{fig:7}). Thus, the resulting influence on the azimuthal temperature distribution at the iris is investigated in the following section.
For this purpose, PTFE adhesive tape (3M\texttrademark{} PTFE Film Electrical Tape 60) was applied directly to the niobium surface in the iris region, as shown in Fig.~\ref{fig:18}. During a 20-minute rf heating experiment using the $8\pi/9$~mode, the infrared camera was mounted at a fixed distance of 20~cm from the cavity and directed towards the iris region on the pickup side. The camera image on the left shows part of the cavity interior, with tape strips appearing as bright regions while rf power is applied. Seven circular measurement positions were defined on the tape strips close to the iris and labeled El1--El7 in clockwise order. For orientation, El1 and El7 are highlighted more prominently. The black dashed line indicates the azimuthal plane corresponding to the slots in the external stiffening rings. For comparison, a partially sectioned CAD model is shown, providing a view of the inner surface of the first cell and its iris region. The corresponding section plane is highlighted in red. The tape was calibrated in situ during the $\pi$-mode measurement (see Fig.~\ref{fig:9}), yielding an emissivity of $\varepsilon = 0.97$. 
\begin{table}[b]
    \centering
    \caption{Average temperatures at the iris measurement positions
    after 20 minutes of rf heating using the $8\pi/9$~mode.}
    \label{tab:3}
    \begin{tabular}{c|ccccccc}
        Position & El1 & El2 & El3 & El4 & El5 & El6 & El7 \\
        \hline
        $T$ ($^\circ$C) & 52.0 & 50.0 & 49.1 & 49.6 & 50.7 & 50.9 & 49.9 \\
    \end{tabular}
\end{table}

For this calibration, square pieces of PTFE tape were attached to the uncoated surface of the central cell near Li5. Their measured temperatures were compared with those of the surrounding graphite-coated surface, following a procedure similar to the initial emissivity calibration. The $8\pi/9$~mode was specifically selected because it exhibits the highest rf power dissipation in the end cells, which are readily accessible for infrared measurements.

The azimuthal temperature variation and the location of the hottest region can thus be estimated, providing a basis for assessing the temperature distributions of other modes within the same dipole passband especially for the $\pi$~mode. Experimental observations support this approach, as the azimuthal positions of the hot spots at the equator coincide for both the $\pi$ and $8\pi/9$~modes, as shown in Figs.~\ref{fig:9} and \ref{fig:11}. In both cases, the hot spots follow the cavity rotation while maintaining their relative positions on the cavity surface. After 20 minutes of rf heating at the applied power level, the average temperatures at the seven measurement positions were determined, as summarized in Table~\ref{tab:3}. The maximum temperature difference among these positions was 2.9~K, with the lowest temperature measured at El3 and the highest at El1, located approximately $90^\circ$ azimuthally from each other.

The CST simulations shown that the maxima of the magnetic energy density in the iris region are shifted azimuthally by $90^\circ$ relative to the corresponding maxima at the equator, as illustrated in Fig.~\ref{fig:7}. Consistent with this distribution, El3 which exhibits the lowest temperature, is located in approximately the same azimuthal plane as the temperature maxima measured externally at the equator, while El1 with the highest temperature, is positioned approximately $90^\circ$ away. Near the slot position of the stiffening rings, a temperature difference of 2.0~K was measured between El1 and El2, located above and below the dashed line in Fig.~\ref{fig:13} or \ref{fig:15}, respectively. This enables temperature measurements performed externally through the slot in the stiffening rings to be associated with the corresponding azimuthal positions in the iris region. The measurement was performed under conditions expected to produce particularly large local temperature variations, as the $8\pi/9$ ~mode exhibits a high local magnetic energy density in the outer cells, resulting in concentrated rf power dissipation in their iris regions. The observed temperature differences can therefore serve as a conservative reference for assessing the azimuthal temperature inhomogeneity of other modes within the same dipole passband.

Applying these findings to the combined $\pi$-mode measurement to estimate the overall temperature spread across the cavity, and assuming that the maximum iris temperature may be higher than the temperature measured at the corresponding position in Li~10, an additional 1~K should be considered in the overall temperature spread of 12.6~K. This accounts for the fact that the measurement was taken close to, but not exactly at, the location of the maximum iris temperature. This results in a conservative estimate of the maximum temperature spread of 13.6~K across the entire cavity at the applied rf heating power.

\section{Summary and conclusion}
In exploring the possibilities of rf heating of SRF cavities, with the long-term goal of enabling in-situ baking within cryomodules, this work investigates HOM heating using the installed HOM couplers of a TESLA 9-cell cavity. Based on their suitable coupling, the four highest-frequency modes of the first TE-like dipole passband were analysed, with the $\pi$~mode emerging as the most promising due to its spatial power distribution. To obtain a more complete picture of the surface temperature distribution using thermography, the tests were performed on an open-flanged cavity under atmospheric conditions, with a graphite-coated outer surface. 
Since the heat-loss mechanisms differ fundamentally from those in a cryomodule configuration, absolute temperatures are not the subject of this investigation.

With an rf power of ($177 \pm 5$)~W coupled into the cavity via the HOM coupler (R) port , resulting in a dissipated rf power of ($140.0 \pm 7.5)$~W, the overall temperature spread across the entire cavity surface was conservatively estimated to be $\Delta T = 13.6$~K. It should be noted that the cavity used in this study, as measured in Ref.~\cite{wk7y-8n6d}, does not exhibit a flat field distribution in the fundamental $\pi$~mode. This condition also applies to higher-order $\pi$~modes. Additional off-axis dipole measurements to characterize the field distribution were not performed in this study.
A direct consequence of an asymmetric rf power deposition towards one side of the cavity, similar to the observations reported in Ref.~\cite{wk7y-8n6d} is observed. This could explain the observed asymmetric heating of the cavity towards the TTF / HOM (R) coupler side. Ensuring a better field flatness would bennefit the overall temperature spread.

Although initially counterintuitive, the observation that the central cells remain cooler in the $\pi$~mode could be beneficial. If the cavity is connected to beam pipes at both ends, additional heat will be conducted away. A higher rf power deposition in the end cells may therefore help compensate for these additional heat losses and contribute to a more uniform temperature distribution.

The $8\pi/9$, $7\pi/9$, and $6\pi/9$ modes also exhibit progressively decreasing but still good coupling characteristics. The HOM coupler (R) was subjected to average rf powers of 190, 201.5, and 193 W $\pm 5$~W, respectively, for 20 min without any noticeable abnormalities during operation or subsequent visual inspection. However, no further examination, including a vacuum leak test, has been performed to date. Due to their inhomogeneous power distribution among the cells, these modes are unsuitable for single-mode heating. However, they are potential candidates for an alternating mode-switching approach, similar to that proposed in previous work \cite{wk7y-8n6d}.

The dipole characteristics of this passband resulted in maximum azimuthal temperature differences along the equator of 4~K for the $\pi$~mode and 4.5~K for the $8\pi/9$~mode at the applied rf power levels.

The transferability of the measured temperature distribution to elevated temperatures under vacuum conditions and with lower surface emissivity is not demonstrated in this paper. Preliminary analytical investigations suggest that no major changes in the temperature distribution are expected under these modified boundary conditions. This topic currently under investigation and will be reported in future work. The present study is therefore limited to the experimental results obtained under the investigated conditions.

Furthermore, additional experiments under conditions as close as possible to those of the intended application are necessary to evaluate the results regarding temperature homogeneity and the maximum achievable temperatures with the installed HOM couplers. In addition, the power handling capability of the HOM couplers in the absence of convective cooling must be determined.

\bibliographystyle{apsrev4-2-titles}
\bibliography{sample_1_}
\end{document}